\documentclass{article}[12pt]

\usepackage{times,xcolor,amsmath,amssymb,epsfig,graphics,graphicx}
\newcommand{\be}{\begin{eqnarray}}
\newcommand{\ee}{\end{eqnarray}}

\begin{document}

\begin{center}
{\Large \bf De Sitter Special Relativity as a Possible Reason for Conformal Symmetry and Confinement in QCD }
\end{center}

\vspace{0.02cm}
\begin{center}
Mariana  Kirchbach$^a $ and Cliffor \ B.\ Compean$^b$\\
$^a$Instituto de F{\'{i}}sica, UASLP,
Av. Manuel Nava 6, Zona Universitaria,\\
San Luis Potos{\'{i}}, S.L.P. 78290, M\'exico\\
$^b$ Facultad de Ciencias, UASLP,\\
 Av. Chapultepec 1570, Privadas del Pedregal,\\
 San Luis Potos{\'{i}}, S.L.P. 78295, M\'exico\\
 E-mails:  mariana@ifisica.uaslp.mx ({\it Corresponding author}), 
cliffor7@gmail.com
\end{center}

\abstract{Conformal symmetry and color confinement in the infrared  regime of QCD  are interpreted by means of a conjectured  deSitter $dS_4$ geometry  of the internal space-time of hadrons, 
an assumption inspired by the hypothesis on deSitter special relativity. Within such a scenario, the  interactions involving the virtual gluon- and constituent quark 
degrees of freedom of hadrons are deduced from the  Green functions of Laplace operators on the  $dS_4$ geodesics. Then the conformal symmetry of QCD emerges as a direct consequence of the conformal symmetry of the $dS_4$ space-time, while the color confinement, understood as colorlessness of hadrons, appears as a consequence of the inevitable  charge neutrality of the unique closed space-like  manifold,  the three dimensional hyper-sphere $S^3$, on whose geodesics the hadron's constituents are conjectured to reside when near rest frame. 
 Mesons  are now modelled  as quarkish color-anticolor dipoles,  whose  free quantum motions on the aforementioned 
$S^3$ geodesics are perturbed  by a  potential generated by a  gluon--anti-gluon color dipole.  The potential predicted presents itself as the color charge analogue to the ``curved'' Coulomb potential, i.e. to the electric potential that defines a consistent electrostatic theory  on a hyper-spherical surface.
The advantage of this method is that it allows to establish a direct relationship of the  potential parameters to the 
fundamental constants of QCD. We apply the model to the description of the spectra of the $a_1$ and $f_1$ mesons, 
and the pion electric charge form factor, finding fair agreement with data.
} 

\vspace{0.3cm}
\begin{flushleft}
PACS: 03.30.+p (Special relativity), 11.30.-j (Symmetry in theory of fields and particles), 12.38.Aw (Quark confinement)\\
Key words: deSitter special relativity,  conformal symmetry,  color confinement, unflavored mesons, strong coupling constant
\end{flushleft}

\begin{flushright}
``Think geometrically, prove algebraically'',\\
 John Torrence Tate Jr.\\
``...verify experimentally!''\\
Galileo
\end{flushright}

\begin{flushleft}
Dedicated to the 60th birthday of Prof.\ Valeri V.  Dvoeglazov.\\
Our wish, an  easy going journey through his successful life, \\
full of  valuable goals, achievements, and memories.
\end{flushleft}

\section{Introduction}
Theoretical models in physics are frequently developed with the  aim to explain particular observational data. Such approaches, termed to as ``bottom-up'',
obviously depend on the data set chosen as the point of departure. It quite might happen that at different energy scales, different models can be adequate. For example,
at low-,  up to pretty high energies, the electron charge in the Quantum Electrodynamics (QED) can be considered as a constant, but in very strong electromagnetic fields, generated in the interiors of stars, and in collisions, it can become dependent on the transferred momentum. Particles interacting via the 
so called ``strong'' interactions, the hadrons,  are described in analogy to QED by a properly designed fundamental gauge theory, the Quantum Chromodynamics (QCD), named in this way in reference to the hypothesis that the hadron constituents, the quarks, are endowed with three different degrees of freedom, conditionally  termed to  as  red-, blue-, and green ``color charges''.  All
strong interacting systems observed so far have been found to be color-neutral, a phenomenon known as color-non-observability, or, color confinement.
The phenomenon has been built into the fundamental gauge theory of strong interactions through a special choice of the gauge group, $SU(3)_c$,  which first prescribes the number of  ``colored'' gauge fields, the gluons, to be eight, and then ``locks'' all the  color degrees of freedom, through limiting the irreducible representations to those of  the factor group with respect to the central charges,  $SU(3)_c/Z_3$.
 The space-time wave equations following from QCD  are  differential equations of high order and difficult to tackle. For the immediate analyzes of the rapidly accumulating data on hadrons concerning  their masses, decay modes, collision probabilities (scattering cross sections), compiled in \cite{PART}, a model has been developed, the  ``constituent'' quark model which is based on second order differential equations of the Sturm-Liouville type, equivalent to employing potentials in the quantum mechanical wave equations of  Schr\"odinger-, Klein-Gordon-, or Dirac. Historically, as point of departure for the traditional quark model \cite{Rayzuddin}  have served  the hadrons of the lowest masses, the pseudo-scalar  mesons $\eta$, $\pi$ (pion), and $K$ (kaon), on the one side, and the
spin-$1/2$  baryons, the nucleon $N$, the $\Sigma$, and $\Xi$ hyperons, as well as their spin-$3/2$ excitations $\Delta$, $\Sigma^*$, and $\Xi^*$, supplemented by the  $\Omega$-Hyperon, on the other. The potentials of conventional use employed so far  within this framework  have been predominantly power 
functions of relative distances. As a rule, the quark models can not be directly linked to the gauge theory of strong interactions, possibly with the exception of the MIT bag model \cite{MIT}, and versions of it \cite{Kharzeev}. To the best of our knowledge, none of the conventional constituent quark models has provided hints on possible reasons behind the confinement, nor  enabled extraction from data on spectra of the value of the fundamental QCD coupling. In a recent work of ours \cite{EPJA16}, the choice of the data set to be explained by a ``bottom-up'' quark model has been shifted from the lowest to the highest  masses of the unflavored mesons which are of the order of 2300 MeV. The phenomenon to be explained there  referred to the observed striking hydrogen-like degeneracies between states distinct through their spins but of same isospin (number of states with different electric charges), and same $CP$ quantum numbers (the product of spatial $(P)$, and charge conjugation ($C$) parities). The explanation suggested  was based on  two fundamental principles. We assumed validity of,

\begin{itemize}

\item deSitter $dS_4$ special relativity \cite{Pereira}, 

\item  interactions defined by Laplacians on geometric manifolds \cite{Kelogg}. 

\end{itemize}

According to the first principle, 
\begin{itemize}
\item the space-time in which the  material bodies, in our case the hadrons, are propagating in real time, and that corresponds to the  ``external space-time'', is described by means of all
the events placed in the interiors of all Minkowski's light cones,  constructed for anyone of the  local observers on the four-dimensional $dS_4$ 
hyperboloid of one shell, ${\mathbf H}_1^4$, via its intersections  by 4D planes passing parallel to the symmetry axis and containing the observer's location,

\item the space-time in which the virtual bodies, in our case the quark and gluon constituents of the hadrons, can interact instantaneously, and  
that corresponds to the  ``internal space-time'',
is described by means of all the events placed outside of the aforementioned light cones.     
\end{itemize}
According to the second principle,  fundamental interactions are defined by  Green functions of Laplace operators on manifolds, the latter being hypothesized by us to be suitably chosen $dS_4$ geodesics.  

\begin{quote}
The principle advantage of fixing the geometry of the internal space-time of hadrons to the $dS_4$ hyperboloid is that among its rich geodesic structure 
there exists in particular one geodesic space that  inevitably and necessarily describes a confinement phenomenon akin to the color charge neutrality in QCD. This is the unique closed space-like hyper-spherical manifold, $S^3$, located at the ``waist'' of the hyperboloid.

\end{quote}
Indeed, it can be shown that the Gauss theorem and the superposition principle allow only charge neutral configurations to exist on such a surface, the lowest configurations being charge dipoles. Moreover, free quantum motions of quark-anti-quark dipoles on $S^3$, when perturbed by the potential generated by a gluon-anti-gluon dipole, give rise to an instantaneous interaction  capable of explaining a large amount of data on the reported meson excitations \cite{EPJA16}, \cite{Addendum2016}. To be specific,  the aforementioned interaction presents itself as the color charge analogue to the ``curved'' Coulomb potential on $S^3$,
given by \cite{Barut}
\begin{eqnarray}
V^{S^3}_{C}=\frac{\ell (\ell +1)}{\sin^2\frac{\stackrel{\frown}{r}}{R}} -\alpha Z \cot \frac{\stackrel{\frown}{r}}{R}\stackrel{R\to\infty}{\longrightarrow}
V_C^{E_3}=
R^2\frac{\ell(\ell+1)}{r^2}-R\frac{\alpha Z}{r},
\end{eqnarray}
where $R$ is the $S^3$ hyper-radius, $\stackrel{\frown}{r}$ denotes the arc of the geodesic (to be specified below) while $V_C^{E_3}$ is the standard Coulomb potential in flat three dimensional Euclidean space, $E_3$, the limiting $V_C^{S^3}$ case for  $R\to \infty$. Then, the strong interaction under discussion is obtained from $V^{S^3}_C$ by replacing the product, $\alpha Z$,  of the fundamental constant $\alpha$ in QED, and the charge number, $Z$, by $\alpha_sN_c$, the product of the strong coupling $\alpha_s$ in QCD and the number of color charges, $N_c$.
The strong interaction potential derived in this way allows one to put the description of mesons, the simplest (two-body)  composite system in QCD,  at comparable footing with that of the H Atom, the simplest (two-body) system  in QED.      
In fitting by the above potential  data on the observed degeneracy patterns of the unflavored meson spectra, predictions for $\alpha_s$ at different meson masses  could be obtained in \cite{Addendum2016} that matched values reported by 
other sources. In \cite{NPhA18} the method under discussion has been further 
successfully tested in the evaluation of the nucleon electromagnetic form-factors.  \\

\noindent
In the present study we first briefly review the proposal of \cite{EPJA16}, \cite{Addendum2016} and then put it at work in the description of the previously not considered  $a_1$ and $f_1$ meson spectra, and also in the explanation of the  pion electric charge form factor, finding very satisfactory agreement with data.

The article is structured as follows. In the next section we highlight the idea of \cite{Pereira} on the extension of  Einstein's special relativity to deSitter relativity. In section 3 we recall the elements of the algebraic description of the deSitter geometry, which allows us to consider there free quantum motions on the unique closed hyper-spherical space-like manifold, as well as on  open hyperbolic time-like geodesics.
In section 4 we transform the aforementioned free motions into one-dimensional quantum mechanical wave equations with the respective trigonometric Scarf well-, and the hyperbolic P\"oschl-Teller barrier potentials. There, we  present a scenario for a duality between the descriptions of hadrons as states  bound  within the Scarf potential, on the one side, and as resonances transmitted through the P\"oschl-Teller  barrier, on the other, two methods  suited for handling particle physics phenomena  in low- and high-energy regimes. In section 5 we explore consequences of the innate charge neutrality  of the hyper-sphere on the internal structure of systems on this space and review the prediction of \cite{EPJA16}  on a confinement phenomenon. In section 6 we associate the predicted confinement phenomenon with the experimentally detected  color confinement in QCD and analyze data on the $f_1$ and $a_1$ mesons, together with the pion electric charge form factor,
 before closing by a short Conclusion section.

\section{The  Geometric Ansatz. Immersing  Minkowski- into deSitter space-time }
All theoretical descriptions of processes in the micro-world have to  obey the theory of Special Relativity
which requires the values of the physical observables to be same in all reference frames related to each other by Lorentz transformations,
known to conserve the ``intervals'',  $s^2$, in a Minkowskian space-time, ${\mathcal M}_{1+3}$,  of one time-like ($ct$)
and three space-like (${\mathbf r}$)  dimensions according to,
\begin{eqnarray}
{\mathcal M}_{1+3}:\quad c^2(t_1-t_2)^2- (x_1-x_2)^2+(y_1-y_2)^2+(z_1-z_2)^2=s^2, &&s^2\stackrel{\geq}{<} 0 .
\label{Mink_sign}
\end{eqnarray}
Here, $x_i$, $y_i$, and $z_i $ (with $i=1,2$) are the  Cartesian position coordinates of two bodies in the ordinary three-dimensional Euclidean space, $E_3$, 
while $t_1$ and $t_2$ are in turn their time coordinates, and $c$ is the speed of light.

\begin{figure}
\centering
{\includegraphics[width=6.95cm]{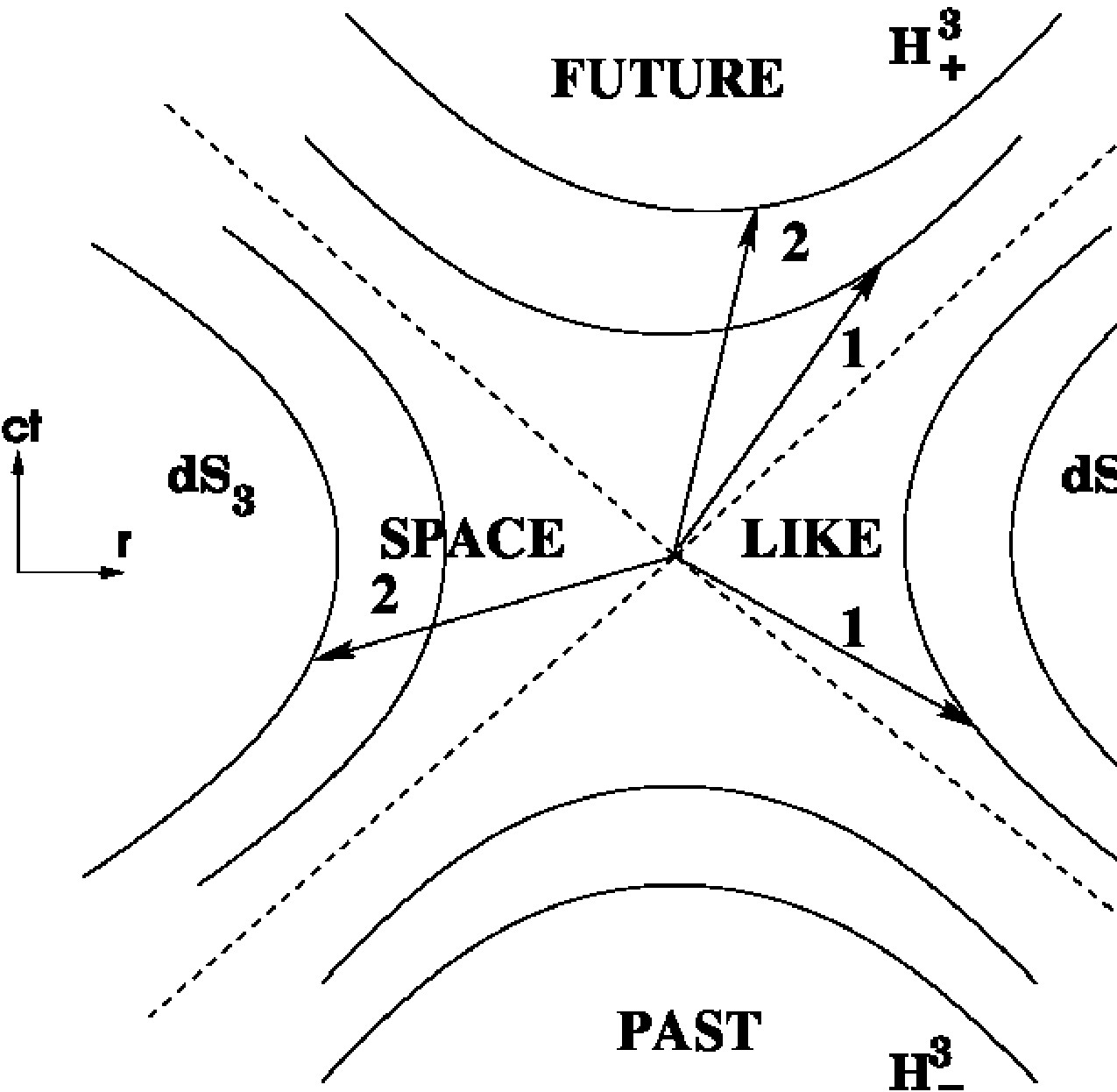}}
\caption{Schematic presentation of  Minkowski's space time,  reduced to a plane.  Two examples of moved reference frames are given and their axes  labeled by  ``1'', and ``2''.  The axes of a frame in motion do not remain mutually orthogonal, the angles between them being dependent on the velocity. A fixed interval, $s^2>0$, defines  a 
three-dimensional hyperboloid of a vertical time-like symmetry axis and of two-sheets represented by the  hyperbola of two branches on the figure, marked by ${\mathbf H}_-^3$ and ${\mathbf H}_+^3$, respectively, or, of one sheet, a so called deSitter $dS_3$ space-time, for $s^2<0$. The dashed lines mark the null-hyperboloid (light cone) corresponding to $s^2=0$. }
\end{figure}
The intervals in (\ref{Mink_sign}) for $s^2>0$ can be viewed as locuses  of points lying on three dimensional hyperboloids of two sheets, denoted by ${\mathbf H}_\pm ^3$,
encapsulated  inside  the ``null hyperboloid'' , $c^2t^2-{\mathbf r}^2=0$, a cone  also termed to as the ``Light Cone'' ,  
as schematically illustrated  in Fig.~1. The null-intervals  remain invariant under five  more transformations termed to as ``conformal'', the resulting symmetry being the ``kinematic conformal symmetry''  \cite{Slava}. The conformal symmetry is associated with all the transformations which 
leave the interval, $ds^2=g_{\mu\nu}(x) dx^\mu dx^\nu$, invariant,  modulo a multiplicative ``conformal factor'' of  $e^{\omega (x)}$. 

In Special Relativity it is argued that the hyperboloids inside the upper part of the cone describe the  causal ``Future'' while those of the bottom part describe the  causal ``Past''.  The meaning of ``causal'' is  that  events in these regions describe material bodies moving with subluminal velocities, because their
time-orderings are preserved by all Lorentz transformations and thus allow one  to define the arrow of time in this domain.
In contrast, the surrounding hyperboloids of one sheet represent the  acausal region, where no arrow of time can be defined because here Lorentz transformations can change the time-orderings of the events. Such events can be used only in the description of virtual degrees of freedom of the material bodies, such as their constituents, among them  the nucleons in a nucleus, the quarks and gluons  in hadrons etc.

Measurements in physics can be carried out only on material bodies, as  described by means of events located in the causal region,  the processes being termed to as  ``real'', or ``on mass-shell'', in reference to the definition of the invariant mass, $M$, by means of the so called ``on-mass shell'' relation, $E^2-{\mathbf p}^2c^2=M^2c^4$, between the energy, $E$, and linear momentum, ${\mathbf p}$.
Instead, the constituents of the material bodies and their instantaneous interactions are exclusively  described by means of events located  in the domain outside the light cone. Notice that the causal and acausal regions are disconnected and  described by means of independent coordinates, a reason for which transformations and  operators from the two regions are commuting with each other. \\

\noindent
Thinking now geometrically, one can ask the question, posed in \cite{Pereira}, on the type of physics one could encounter by imagining  the light cone and the enclosed ${\mathbf H}_\pm^3$ hyperboloids 
immersed locally into a four-dimensional hyperboloid of one sheet as schematically represented in Fig.~2 . Stated differently, one could entertain the idea of  extending  the inaccessible to direct measurements space-like region of the Minkwski space-time by one extra space-like (infinite) dimension, $x_4$, and requiring outside the causal region invariance of the larger intervals, 
\begin{equation}
dS_4:\quad c^2t^2  -{\mathbf r}^2 -x_4^2={\mathcal S}^2,\quad {\mathcal S}^2=-R^2<0.
\label{dS4}
\end{equation} 
Here, $R$ is a new real length parameter.
A space-time defined in this way is known under the name of  ``four-dimensional deSitter space-time'', abbreviated, $dS_4$.
It represents a four dimensional hyperboloid of one sheet, denoted by ${\mathbf H}^4_1$,  with the time line  as a symmetry axis.
Within this set up, causal hyperboloids  in Einstein's special relativity correspond to so called causal patches on $dS_4$ obtained through 
intersections (slicings) of ${\mathbf H}_1^4$   by four-dimensional planes parallel to the time axis as depicted in Figure 2.
The interval in (\ref{dS4}) is conserved besides under Lorentz transformations also under transformations which ensure the transitivity of the $dS_4$ space-time and combine in a particular way
 translations with four of the five conformal transformations, the so called special conformal transformations, a reason for which $dS_4$ features conformal symmetry \cite{Pereira}. In the following we will explore consequences of this idea.

\begin{figure}
\centering
{\includegraphics[width=2.5cm]{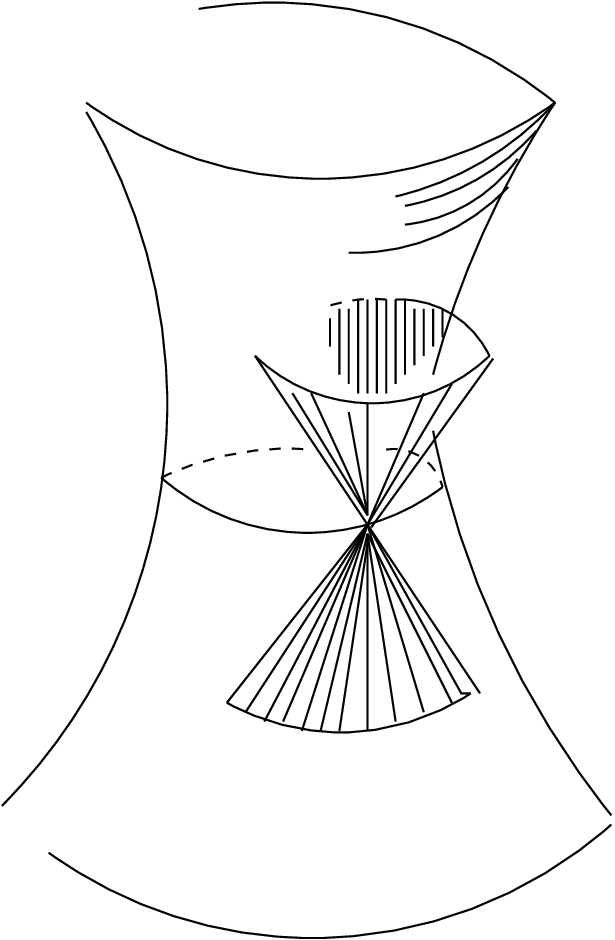}}
\caption{Schematic illustrative presentation of a causal Minkowski light cone relevant to a local observer placed at the ``waist'' of the $dS_4$ hyperboloid, for concreteness. }
\end{figure}

\section{Building up the algebraic description  of the $dS_4$ geometry }

In this section, which partially follows our previous work \cite{EPJA16},  
we recall once again the usefulness of the $dS_4$ geometry  for 
quantum physics. For this purpose we start building up  the algebraic description of quantum motions there, beginning with  free motions.
Free quantum motions on any space-time are described by the eigenmodes of the kinetic energy operator there, 
given by the respective  Laplacian.
On $dS_4$, parameterized in global coordinates as \cite{Coreans},
\begin{eqnarray}
x^0&=&R \sinh\rho, \quad \rho\in (-\infty,+\infty),\nonumber\\
x^4&=&R \cosh\rho \sin\chi, \quad \chi\in\left[-\frac{\pi}{2},+\frac{\pi}{2} \right],\nonumber\\
x^1&=&R \cosh\rho \cos\chi\sin\theta, \quad  \theta\in\left[-\frac{\pi}{2},+\frac{\pi}{2} \right],  \nonumber\\
x^2&=&R\cosh\rho \cos\chi\cos\theta \sin\varphi, \quad \varphi\in \left[0,2\pi \right],\nonumber\\
x^3&=&R\cosh\rho \cos\chi\cos\theta \cos\varphi,
\label{chart}
\end{eqnarray} 
one encounters  the Laplacian, denoted by $\Delta_{dS_4}(\rho,\chi,\theta,\varphi)$ as,
\begin{eqnarray}
\Delta_{dS_{4}}(\rho,\chi,\theta,\varphi ) &=&\frac{1}{R^2\cosh^3\rho }
\frac{\partial}{\partial \rho}\cosh^3\rho \frac{\partial}{\partial \rho}
+\frac{ {\mathcal K}^2(\chi,\theta,\varphi)}{R^2 \cosh^2\rho}.\label{LBLTR}
\end{eqnarray}
The  ${\mathcal K}^2(\chi,\theta,\varphi)$ part of $\Delta_{dS_4}(\rho,\chi,\theta,\varphi)$ in (\ref{LBLTR}) stands for the squared four dimensional ($4D$) angular momentum operator,
\begin{eqnarray}
{\mathcal K}^2(\chi,\theta,\varphi)=-R^2\Delta_{S^3}(\chi,\theta,\varphi)&=&
-\frac{1}{\cos^2\chi} \frac{\partial }{\partial \chi}\cos^2\chi \frac{\partial}{\partial \chi }
+\frac{
{\mathbf L}^2(\theta,\varphi)}{\cos^2\chi},
\label{LB_S3}
\end{eqnarray}
where $\Delta_{S^3}(\chi,\theta,\varphi)$ is the Laplace operator on the unique  closed space-like manifold \cite{Prestley} on $dS_4$ which is a three dimensional hyper-sphere, $S^3$,  of (hyper)radius $R$.
The  $\Delta_{S^3}(\chi,\theta,\varphi)$ eigenmodes are determined  by the hyper-spherical harmonics,   $Y_{K\ell m} (\chi,\theta,\varphi)$, as
\begin{eqnarray}
-\hbar^2c^2\Delta_{S^3}(\chi,\theta,\varphi) Y_{K\ell m} (\chi,\theta,\varphi) &=& \frac{\hbar^2c^2}{R^2} K(K+2)Y_{K\ell m }(\chi, \theta,\varphi),
\quad |m|\in [0,\ell], \label{RosMor_pre}
\end{eqnarray}
explicitly given by,
\begin{eqnarray}
Y_{K\ell m}(\chi,\theta,\varphi)&=&{\mathcal S}_{n\ell}(\chi)Y_\ell^m(\theta,\varphi), \label{4Dharm}\\
{\mathcal S}_{n\ell}(\chi) &=&\cos^\ell \chi {\mathcal G}^{\ell +1}_n(\sin\chi),\quad n=K-\ell.
\label{n4}
\end{eqnarray}
Here,  ${\mathcal G}^{\ell +1}_n(\sin\chi)$, and $Y_\ell^m(\theta,\varphi)$ 
in turn denote  the Gegenbauer polynomials, and the ordinary spherical harmonics. They describe rigid ``dumbbells''  (  `` rotators'')  in four Euclidean dimensions, freely hovering around their  mass centers, with the ends tracing  $S^3$ great circles.

The free waves of the $dS_4$ Laplacian in (\ref{LBLTR}) are then described in terms of 5D pseudo-spherical harmonics,
de\-no\-ted by $Y_{\bar K K \ell m }(\rho,\chi,\theta,\varphi)$, the solutions to,
\begin{eqnarray}
-{\hbar^2 c^2}\Delta_{dS_{4}} (\rho,\chi,\theta,\varphi) Y_{{\bar K} K \ell m}(\rho, \chi, \theta,\varphi)&=& -\frac{\hbar^2c^2}{R^2}{\bar K} ({\bar K} +3) Y_{{\bar K} K\ell m}(\rho, \chi, \theta,\varphi),
\label{5waves}
\end{eqnarray}
 and are defined as,
\begin{eqnarray}
Y_{{\bar K} K\ell m}(\rho, \chi,\theta,\varphi) &=&\phi_{\bar K}(\rho) Y_{K\ell m}(\chi,\theta,\varphi),\quad \bar K, K=0,1,2...,\quad \ell =0,1,2,..., K,
\label{Y_eoneplus4}
\end{eqnarray}
where ${\bar K}$ and $K$ can take all non-negative integer values.

Here, the $\phi_{\bar K}(\rho)$ functions express in terms of Jacobi's polynomials, $P^{a b}_{n_r}(x)$,
 of equal real parameters,  and a pure imaginary argument according to,
\begin{eqnarray}
\phi_{\bar K}(\rho)&=&\cosh^{-\frac{a}{2}-\frac{3}{2}}\rho P_{n_r}^{-a-\frac{1}{2}, -a-\frac{1}{2}}(i\sinh\rho ),\label{wafus}\\
a&=&\frac{1}{2} +\left( {K} +1\right),
\label{ad}\\
P_{n_r}^{-a-\frac{1}{2}, -a-\frac{1}{2}}(i\sinh \rho)&=&\frac{\left(-a+\frac{1}{2}\right)_n} {n_r!}{_2}F_1\left(-n_r, n_r-2a;-a+\frac{1}{2};\frac{1-i\sinh\rho }{2} \right).
\label{nr}
\end{eqnarray}
Furthermore,   $_2F_1$ is the hyper-geometric function, $(...)_t$ is the Pochhammer symbol,  $\rho$ is the arc of an open time like hyperbolic geodesic on $dS_4$, and $n_r$ denotes the polynomial degree.

In summary, the equations (\ref{5waves}) and (\ref{RosMor_pre}) describe free quantum motions along the respective open time-like and 
closed space-like geodesics on $dS_4$
in the sense that the arguments of the wave functions $\phi_{\bar K}(\rho)$, and ${\mathcal S}_{\ell n}(\chi)$ in the respective equations 
(\ref{wafus}) and (\ref{n4}), change along the arcs, $R\rho$, and $R\chi$,  of great circles on  such geodesics. The open time-like geodesics considered here are hyperbolic deSitter space-times of one less dimension, i.e. $dS_3$ space-times, while the unique closed space-like geodesic is a three dimensional (hyper)spherical surface $S^3$ located at the ``waist'' of the $dS_4$ hyperboloid.


\section{From free quantum motions on $dS_4$ to one-dimensional stationary wave equations with centrifugal  potentials  }

The free quantum motions on the open time-like--, and the closed space-like $dS_4$ geodesics can be transformed in their turn into stationary quantum mechanical wave equations  with the one-dimensional hyperbolic  P\"oschl-Teller potential, $V_{\mbox{ PT}}(\rho)$,   and the trigonometric Scarf potential,
$V_{\mbox{Sc}}(\chi)$. 

\subsection{Quantum motions on open time-like geodesics and the hyperbolic P\"oschl-Teller barrier}
To prove the aforementioned statement, the following variable change in (\ref{wafus}) has to be performed,
\begin{equation}
F_{\bar K n_r}(\rho)=\cosh^{\frac{3}{2}}\rho\, \phi_{\bar K}(\rho), \quad {\bar K}=K-n_r,
\end{equation}
in which case the equation  (\ref{5waves}) is similarity transformed toward,

\begin{eqnarray}
-{\hbar^2 c^2}\left[ \cosh^{\frac{3}{2}}\, \rho\,   \Delta_{dS_{4}} (\rho,\chi,\theta,\varphi)\cosh^{-\frac{3}{2}}\rho\right]
&&\left[ \cosh^{\frac{3}{2}}\rho Y_{{\bar K} K \ell m}(\rho, \chi, \theta,\varphi)\right]\nonumber\\
&=& -\frac{\hbar^2c^2}{R^2}{\bar K} ({\bar K} +3)\left[ \cosh^{\frac{3}{2}}\rho Y_{{\bar K} K\ell m}(\rho, \chi, \theta,\varphi)\right].\nonumber\\
\label{paso1}
\end{eqnarray}
Introducing now the notation, ${\mathcal H}_{\mbox{PT}}(\rho,\chi,\theta,\varphi)$, for the  Laplacian,
$\left[ -\hbar^2c^2\Delta_{dS_4}(\rho,\chi,\theta,\varphi)\right]$, upon its  similarity transformation by the $\cos^{\frac{3}{2}}\chi$ function,
\begin{eqnarray}
{\mathcal H}_{\mbox{PT}}(\rho,\chi,\theta,\varphi)&=&\cosh^{\frac{3}{2}}\rho \left[ -{\hbar^2 c^2}\Delta_{dS_{4}} (\rho,\chi,\theta,\varphi) \right]\cosh^{-\frac{3}{2}}\rho,
\label{paso2}
\end{eqnarray}
and making use of  (\ref{LBLTR}) and  (\ref{Y_eoneplus4}), the equation (\ref{paso1})  equivalently rewrites as,
\begin{eqnarray} 
{\mathcal H}_{\mbox{PT}}(\rho,\chi,\theta,\varphi)F_{{\bar K}n_r}(\rho)Y_{K\ell m}(\chi,\theta,\varphi)&=&
\frac{\hbar^2c^2}{R^2} \left[ -\frac{d^2}{d\rho ^2}  +V_{PT}(\rho) +\frac{3^2}{2^2} \right]{\mathbf I}(\chi,\theta,\varphi)\nonumber\\
&\times&  F_{{\bar K} n_r}(\rho)Y_{K\ell m}(\chi,\theta, \varphi)\nonumber\\
&=& -\frac{\hbar^2c^2}{R^2}{\bar K}({\bar K} +3) F_{{\bar K}n_r}(\rho)Y_{K\ell m}(\chi,\theta,\varphi).
\label{PT_pot}
\end{eqnarray}
Here, ${\mathbf I}(\chi,\theta,\varphi)$ is the identity operator in the $S^3(\chi,\theta,\varphi)$ subspace,  $V_{\mbox{PT}}(\rho)$ stands for the P\"oschl-Teller (PT) potential given by,
\begin{eqnarray} 
{\mathcal V}_{\mbox{PT}}(\rho)&=&\left[(K+1)^2 -\frac{1}{4}\right]\mbox{sech}^2\rho, \quad \rho \in (-\infty, +\infty).
\label{PoschTell}
\end{eqnarray}
Notice that the similarity transformation conserves the $ \left[-{\hbar^2 c^2}\Delta_{dS_{4}} (\rho,\chi,\theta,\varphi)\right]$ eigenvalues in (\ref{5waves}).
A way of rephrasing the similarity transformation is to cast it in the form of an intertwinement by the $\cosh^{\frac{3}{2}}\rho$ function of ${\mathcal H}_{\mbox{PT}}(\rho,\chi,\theta,\varphi)$ with the $\left[-{\hbar^2 c^2}\Delta_{dS_{4}} (\rho,\chi,\theta,\varphi)\right]$ Laplacian  according to,
\begin{equation}
-\hbar^2 c^2 \cosh^{\frac{3}{2}}\rho\, \Delta_{dS_{4}} (\rho,\chi,\theta,\varphi)=
 {\mathcal H}_{\mbox{PT}}(\rho,\chi,\theta,\varphi)\, \cosh^{\frac{3}{2}}\rho\, ,
\label{intrtwnd1}
\end{equation}
and to recall that intertwined Hamiltonians are isospectral \cite{Khare}.
Also notice that the PT potential can be equivalently rewritten to,
\begin{equation}
{\mathcal V}_{\mbox{PT}}(\rho)=\left[ (K+1)^2-\frac{1}{4}\right] \frac{1}{\cosh^2\rho}=-\left[ (K+1)^2-\frac{1}{4}\right]
\mbox{tanh}^2\rho +\frac{\hbar^2c^2}{R^2},
\label{Higgs_osc}
\end{equation}
where the potential on the r.h.s. is  known under the name of the ``Higgs oscillator''  on $dS_4$ in reference to the fact that the leading term in the  series expansion of $\tanh^2\rho$ goes as the square, $\rho^2$, of the argument meaning that to leading order the hyperbolic P\"oschl-Teller potential behaves as an oscillator.
In this sense, free motions on the open hyperbolic time-like geodesics on $dS_4$ can be viewed as ``Higgs'' oscillations. 
An  interesting phenomenon that can be related to free quantum motion on the $dS_4$ space is  predicted upon the  complexification, $(K+1)\longrightarrow i(K+1)$, of the parameter defining the potential strength in (\ref{PoschTell}), 
equivalently, upon the analytical continuation of the four-dimensional angular momentum to complex values.
In this case   ${\mathcal V}_{\mbox{PT}}(\rho)$ is transformed into a barrier and one can consider the complex energies of the resonances transmitted through it.
To do so, one has to calculate the  transmission scattering matrix, $T(k)$. A scheme for such a calculation has been presented in detail in  \cite{Cevik}. The expression obtained along the lines of \cite{Cevik} for the case of our interest  reads,
\begin{eqnarray}
T(k)&=&\frac{\mbox{sinh}^2\pi k }{\mbox{cosh}^2\pi k +\mbox{sinh}^2 \pi (K+1)  }.
\label{TrnsmMtrx}
\end{eqnarray}
Then  the poles, which  appear at $k= \left[ \left( K+1\right) +i\left( n_r-\frac{1}{2}\right) \right]$, define the  squared complex energies, $\left({\mathcal E}{}^{(res)}\right)^2$, of the resonances transmitted through the barrier as \cite{EPJA16},
\begin{eqnarray}
\left( {\mathcal E}{}^{(res)}\right)^2=\frac{\hbar^2c^2}{R^2}k^2,\quad
{\mathcal I}m\,\left( {\mathcal E}^{(res)}\right)^2&=& 2\frac{\hbar^2c^2}{R^2} \left(K+1\right)\left(n_r-\frac{1}{2}\right).
\label{Im_E2Res}\\
{\mathcal R}e\,\left( {\mathcal E}^{(res)}\right)^2-c_0&=&\frac{\hbar^2c^2}{R^2}\left(K+ 1\right)^2,\nonumber\\
 c_0 &=& -\frac{\hbar^2c^2}{R^2}\left( n_r-\frac{1}{2}\right)^2 +A^2,
\label{Real_E2Res}
\end{eqnarray}
where $A^2$ is some additive ad hock constant which on long term  could be helpful in adjustments of  the potential parameters to data. Let us interpret the real parts of the squared complex energies,
$\left[{\mathcal R}e\,\left( {\mathcal E}^{(res)}\right)^2-c_0\right]$,
at the poles of $T(k)$,  as a squared invariant mass  $M^2$, i.e. let us  set,
\begin{eqnarray}
M^2&\equiv &{\mathcal R}e \,\,\left( {\mathcal E}^{(res)}\right)^2-c_o=
\frac{\hbar^2 c^2}{R^2}(K+1)^2=\frac{\hbar^2 c^2}{R^2}(n+\ell +1)^2,\quad K=n+\ell.
\label{lin_masses}
\end{eqnarray}
The expression gives rise to linear ``trajectories''  (linear dependencies of the angular momentum on the mass)  of the art,
\begin{eqnarray}
\ell = \alpha (R) M  -n- 1, && \ell=0,1,2,..., \quad \ell +n=K, \quad K=0,1,2,...
\label{My_trjct}
\end{eqnarray} 
with $n$ being the number of nodes in the Gegenbauer polynomials  in (\ref{n4}).

\subsection{Quantum motion on the closed space-like hyper-spherical geodesics and the trigonometric Scarf potential}
In a similar way, through the variable change,
\begin{eqnarray}
\cos\chi Y_{Klm}(\chi,\theta,\varphi)  &=& \cos\chi {\mathcal S}_{n\ell}(\chi)Y_{\ell m}(\theta,\varphi)\nonumber\\
&=&U_{ n\ell }(\chi)Y_\ell^m(\theta,\varphi), \quad  U_{n\ell }(\chi)=
\cos\chi {\mathcal S}_{n\ell}(\chi),
\label{var_chng_1}
\end{eqnarray}
with ${\mathcal S}_{n\ell}(\chi)$ from  (\ref{n4}),
the  equation (\ref{RosMor_pre}) is similarity transformed  toward,
\begin{eqnarray}
-\hbar^2c^2\left[\cos\chi \Delta_{S^3}(\chi,\theta,\varphi) \cos^{-1}\chi \right]&&\left[\cos\chi Y_{K\ell m}(\chi,\theta,\varphi) \right]\nonumber\\
 &=& \frac{\hbar^2c^2}{R^2} K(K+2)\left[\cos\chi Y_{K\ell m }(\chi, \theta,\varphi)\right].
\end{eqnarray}
 In now introducing the notation, ${\mathcal H}_{\mbox{Sc}}(\chi,\theta,\varphi)$,
for the La\-pla\-ci\-an, 
$\left[ -\hbar^2c^2\Delta_{S^3}(\chi,\theta,\varphi)\right]$, upon its similarity transformation by the $\cos\chi$ function,
 \begin{eqnarray}
{\mathcal H}_{\mbox{Sc}}(\chi,\theta,\varphi) &=&-\hbar^2c^2\left[\cos\chi \Delta_{S^3}(\chi,\theta,\varphi) \cos^{-1}\chi \right],
\label{paso11}
\end{eqnarray}
making use of (\ref{4Dharm}), (\ref{n4}), and (\ref{LB_S3}), and upon some algebraic manipulations,  the equation (\ref{paso11}) equivalently rewrites as,

\begin{eqnarray}
H_{\mbox{Sc}}(\chi, \theta,\varphi)U_{\ell n}(\chi)Y_{\ell m}(\theta,\varphi)&=&
\frac{\hbar^2c^2}{R^2}\left( -\frac{{\mathrm d}^2}{{\mathrm d}\chi^2} +
V_{\mbox{Sc}}(\chi) -1 \right)\otimes {\mathbf I}(\theta,\varphi) U_{\ell n}(\chi)Y_{\ell m}(\theta,\varphi)\nonumber\\
& =&\frac{\hbar^2c^2}{R^2}K(K+2) U_{\ell n}(\chi)Y_{\ell m}(\theta,\varphi),
\label{ScarfI}
\end{eqnarray}
where ${\mathbf I}(\theta,\varphi)$ is the identity operator on the $S^2(\theta,\varphi)$ 
subspace of $S^3(\chi,\theta,\varphi)$, and $V_{\mbox{Sc}}(\chi)$ is the trigonometric Scarf potential
\begin{eqnarray}
 V_{\mbox{Sc}}(\chi)&=& \ell(\ell +1)\sec^2\chi,\quad \chi\in \left[-\frac{\pi}{2},+\frac{\pi}{2} \right].
\label{Scraf}
\end{eqnarray}
Notice that the barrier potential in (\ref{PoschTell}) can be obtained from the well potential in (\ref{Scraf}) through complexification of the argument, modulo multiplicative constant.

Same as in the case of the $dS_4$ Laplacian, also the similarity transformation of the $S^3$ Laplacian conserves its 
eigenvalues in (\ref{RosMor_pre}) and  because these eigenvalues are bound from below, they are referred to as the Laplacian's ``spectrum'' .


A  way of rephrasing the similarity transformation in (\ref{paso11}) is to cast it in the form of an intertwinement by the $\cos\chi $ function of ${\mathcal H}_{\mbox{Sc}}(\chi,\theta,\varphi)$ and the $\left[ -\hbar^2 c^2\Delta_{S^3} (\chi,\theta,\varphi)\right]$ Laplacian according to
\begin{equation}
-\hbar^2 c^2\cos \chi\, \Delta_{S^3} (\chi,\theta,\varphi)= {\mathcal H}_{\mbox{Sc}}(\chi,\theta,\varphi)\cos\chi \, ,
\label{intrtwnd2}
\end{equation}
and recall isospectrality of intertwined operators known from the super-symmetric quantum mechanics \cite{Khare}.
The considerations from above show that in terms of one-dimensional wave equations, the free quantum motions along the open time-like, and the closed space-like $dS_4$ geodesics give in turn rise to  the hyperbolic P\"oschl-Teller, $V_{\mbox{PT}}(\rho)$, equivalently, the Higgs oscillator on $dS_4$, and the trigonometric Scarf, $V_{\mbox{Sc}}(\chi)$ potential,  which are  well known from the super-symmetric quantum mechanics \cite{Khare} to be exactly solvable.
More important,  the Hamiltonians ${\mathcal H}_{\mbox{PT}}(\rho,\chi,\theta,\varphi)$ and ${\mathcal H}_{\mbox{Sc}}(\chi,\theta,\varphi)$, in being intertwined with  the respective
$dS_4$ and $S^3$ Laplacians, conserve the symmetry of the free geodesic motions which is  the conformal symmetry, a reason for which the one dimensional potentials in
(\ref{PoschTell})-(\ref{Higgs_osc}),  and (\ref{Scraf}) are said to be conformal.  Such occurred  because the latter potentials correspond to ``centrifugal'' terms on the respective hyperbolic and hyper-spherical manifolds. We now
 rename by  $\left({\mathcal E}{}^{(bound)}\right)^2$ the $H_{\mbox{Sc}}(\chi,\theta,\varphi) $ eigenvalues in (\ref{ScarfI}), i.e. we introduce the notation,
\begin{eqnarray}
\left({\mathcal E}{}^{(bound)}\right)^2= \frac{\hbar^2c^2}{R^2}(K+1)^2,\quad K=n+\ell .
\label{InvMass_ReE2}
\end{eqnarray}
Because the energy depends on the $K$ quantum number alone, and due to the branching rules,
\begin{equation}
\ell \in [0,K], \quad |m|\in [0, \ell],
\label{brnchng}
\end{equation} 
as explained in the equation (\ref{Y_eoneplus4}), one immediately notices that the levels of the trigonometric Scarf potential are $\sum_{\ell =0}^{\ell =K}(2\ell +1)=(K+1)^2$-fold degenerate.
This type of degeneracy is known from the Hydrogen Atom, where it is attributed to the conformal symmetry of the Maxwell equations, or, more general, 
to the conformal symmetry of the Relativistic Electrodynamics \cite{ED_CS}. Therefore, the eigenmodes of the  free quantum motion along the closed $S^3$ hyper-spherical geodesics on $dS_4$ fall into a  conformal spectrum.\\

\subsection{Equality between the masses of the resonances transmitted through the hyperbolic P\"oschl-Teller barrier and the excitation energies of the states bound within the trigonometric Scarf potential}
Comparison of the real parts of the complex masses of the resonances transmitted through the P\"oschl-Teller barrier in (\ref{lin_masses}) with the real excitation energies of the states bound within the trigonometric Scarf potential in (\ref{InvMass_ReE2}) reveals their equality as,
\begin{equation}
M^2={\mathcal R}e \left({\mathcal E}^{(res)}\right)^2=\frac{\hbar ^2c^2}{R^2}(K+1)^2.
\label{bomba}
\end{equation}

Within the geometric context  pursued  through the text,  the duality between the mass-formulas in (\ref{My_trjct}) and 
(\ref{InvMass_ReE2}),  translates into duality between quantum  motions on open hyperbolic time-like, and the closed hyper spherical space-like  geodesics on $dS_4$.

\begin{quote}
In conclusion,  with the aid of the $dS_4$ structure of the domain located outside the causal Minkowski light cones, hypothesized by us to provide the relevant geometry of internal space-time,  and employed in the definition of interactions, a pair of potentials could be encountered with the remarkable properties that the  real energies of the states bound within the Scarf well potential equal the real parts of the complex energies of the resonances transmitted to the P\"oschl-Teller  barrier. Moreover, these energies carry same hydrogen-like degeneracies.
\end{quote}
In this manner, by virtue of the hypothesis on validity of $dS_4$ special relativity \cite{Pereira}, we were able to establish in \cite{EPJA16}
a duality between the free quantum motions on open hyperbolic time-like--, and  closed hyper-spherical space-like geodesics of the $dS_4$ space-time. To the end of the article, the physical significance of such a duality will be revealed. Before, in the next section, a deeper insight into the physics on the closed hyper-spherical surface $S^3$ will be gained.

\section{ Behind the curtains of confinement. Charge neutrality on the hyper-spherical $dS_4$ geodesics and  a conformal color-dipole  confining potential}
The main point of the present section is to explore  consequences of  the fact that  no single charge  can be consistently defined on  spheres \cite{LandLif}.
Indeed, the lines pouring out of a charge located at a particular point on this space
intersect at the antipodal point, creating there  a fictitious opposite charge (see Fig.~3).  In order to satisfy the Gauss theorem and the principle of superposition on $S^3$, fictitious charges have to be avoided and  
replaced by real charges, thus creating  necessarily and inevitably the charge neutrality of the sphere.  In order to see this, we recall that the Gauss theorem  predicts a $\sec^2\chi$ functional form  \cite{PouriaPedram} for the electric field for $\chi\in [-\pi/2,+\pi/2]$.  On the other side,  the superposition principle on $S^3$  predicts same  field to be  the gradient of the  sum of the  pod- and anti-pod single charge potentials. In the following we shall show that the $E$ field of a ``charge-dipole'' (for the time being without specification of the nature of the charge) predicted by the superposition principle on $S^3$ coincides with the corresponding predictions of the Gauss theorem. In this way we shall prove that a consistent ``charge-static'' 
on the hyper-sphere is guaranteed by its charge neutrality.

\begin{figure}
\centering
{\includegraphics[width=2.5cm]{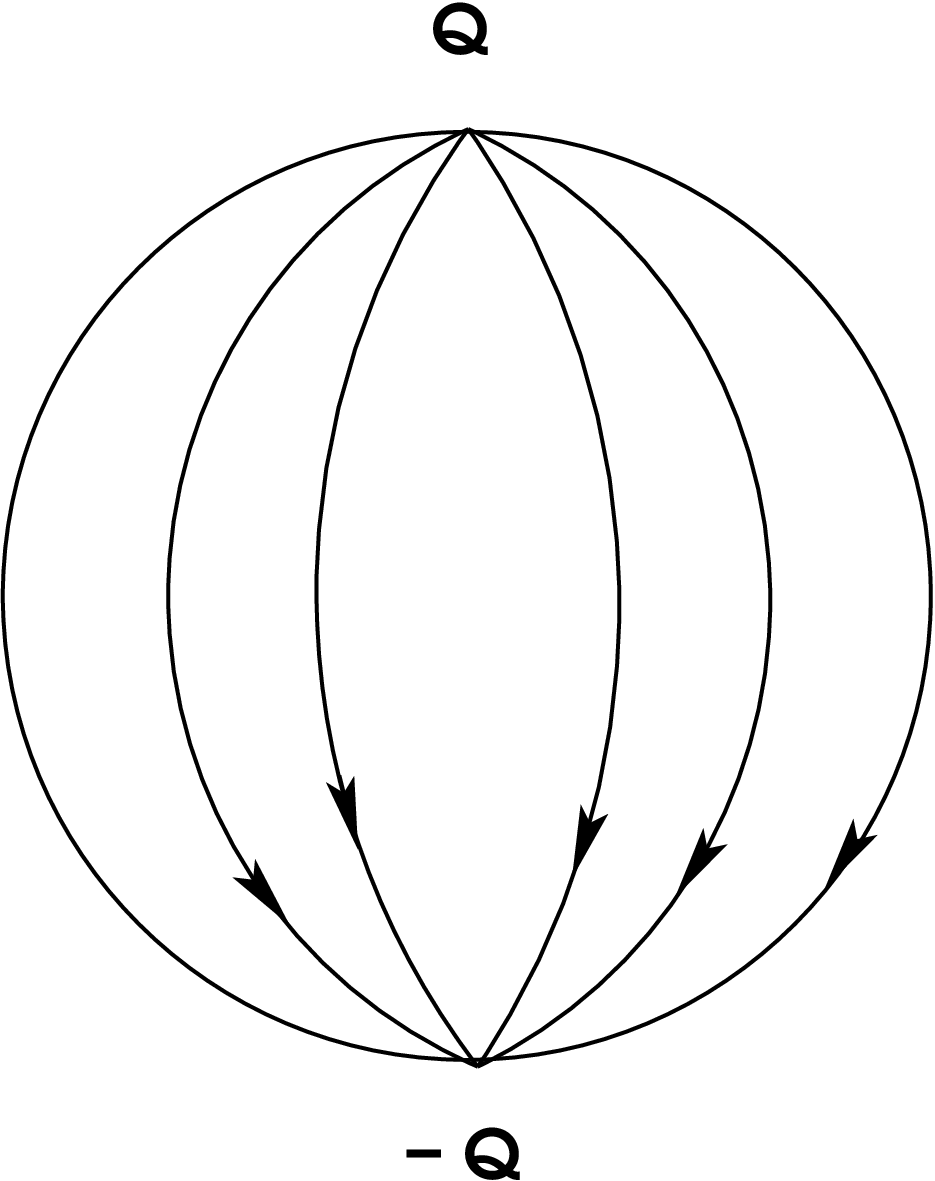}}
\caption{The inevitable charge neutrality of the sphere.  }
\end{figure}
{}For this purpose,
one needs to know the Laplacian on the complete space-time $R^1\otimes S^3$, with $R^1$ standing for the time line, and then to consider it at  constant times.
This can be done within the framework of the so called Radial Quantization Technique which prescribes to first Wick-rotate via the complexification, $ct\to ict$, the
 Minkowski space time of $(1+3)$ dimensions to four Euclidean dimensions, switch to global hyper-spherical coordinates, and introduce ``time'' via the logarithm of the $S^3$ radius \cite{Fubini}. In so doing one finds the following Laplacian,

\begin{eqnarray}
-\left[\frac{\partial ^2}{\partial x_4^2}+ \frac{\partial ^2}{\partial x_1^2}+ \frac{\partial ^2}{\partial x_2^2}
+\frac{\partial ^2}{\partial x_3^2}\right]&\longrightarrow& -\left[\frac{1}{R^3}\frac{\partial }{\partial R}R^3 \frac{\partial }{\partial R}
-\frac{1}{R^2}{\mathcal K}^2(\chi,\theta,\varphi)\right]\label{Wick_Polar}\\
&\stackrel{{R=e^\tau}}{\longrightarrow}&e^{2\tau}\left[ - \frac{\partial^2}{\partial \tau ^2} +\left({\mathcal K}^2(\chi,\theta,\varphi)+1\right) \right].
\label{Conf_Lplc}
\end{eqnarray}
The $\tau$  parameter is known as ``conformal time''.
To build up charge-statics, one has to set $\tau$=const and to calculate the Green function to the remaining piece, the so called conformal Laplacian, $\Delta^1_{S^3}(\chi,\theta,\varphi)$, on $S^3$,  read off from (\ref{Conf_Lplc}) as
\begin{eqnarray}
\Delta^1_{S^3}(\chi,\theta,\varphi)&=&
{\mathcal K}^2(\chi,\theta,\varphi)+1,
\label{conf_LPL}
\end{eqnarray}
where we have chosen $\tau=0$, equivalent to $R=1$.
In order to construct the potentials of a charge ${\mathcal Q}$ placed at, say, the West ``pole'', and of an anti-charge, 
$\left(-{\mathcal Q}\right)$,  at the East `` pole'', one needs to know the respective  Green functions, in turn denoted by  ${\mathcal G}_{-\frac{\pi}{2}}(\chi)$, and ${\mathcal G}_{+\frac{\pi}{2}}(\chi)$, which have been calculated, among others, in \cite{Birgitta}, however for $\chi\in \left[0,\pi\right]$, and need to be shifted to $\chi\in \left[ -\frac{\pi}{2},+\frac{\pi}{2}\right]$.
After this small algebraic manipulation, the potentials,  constructed from the Green's function in the standard way known from potential theory \cite{Kelogg}, emerge as,
\begin{eqnarray}
 V_{-\frac{\pi}{2}}(\chi)={\mathcal Q}{\mathcal G}_{ -\frac{\pi}{2}}(\chi) &=&-\frac{{\mathcal Q}}{4\pi^2}\left(\chi -\frac{3\pi}{2}\right)\tan \chi,
\label{VNP} \\
V_{+\frac{\pi}{2}}(\chi) =-{\mathcal Q}{\mathcal G}_{ +\frac{\pi}{2} }(\chi) &=&-\frac{(-{\mathcal Q})}{4\pi^2}\left(\chi -\frac{\pi}{2} \right)\tan \chi, \quad 
\chi\in\left[-\frac{\pi}{2},+\frac{\pi}{2} \right]. 
\label{VSP}
\end{eqnarray}
In now assuming validity of the superposition principle, one encounters a  Charge Dipole (CD) potential to emerge at a point $\chi$ on $S^3$ according to,
\begin{eqnarray}
V_{CD}(\chi) &=&{\mathcal Q}{\mathcal G}_{-\frac{\pi}{2} }(\chi) -{\mathcal Q}{\mathcal G}_{+\frac{\pi}{2}} (\chi)  =\frac{{\mathcal Q}}{4\pi}\tan \chi. 
\label{Clr_Dpl_pre}
\end{eqnarray}
The electric field to this dipole is obtained in the standard way through differentiation as,
\begin{equation}
E(\chi)=-\frac{\partial }{\partial \chi}V_{CD}(\chi)=-\frac{{\mathcal Q}}{4\pi}\frac{1}{\cos^2\chi}. 
\label{Efield_on_S3}
\end{equation}
On the one side, this  is the precise expression prescribed by   Gauss's theorem \cite{PouriaPedram}, and on the other, one
recognizes in it the functional form of the Scarf well. 
As an important reminder, $V_{CD}(\chi)$ is conformal because the Green functions have the symmetry of the Laplacian, that on its side has the symmetry of the space-time on which it defines the kinetic energy operator. 

One of the advantages of defining non-relativistic potentials in terms of Green functions is that elements of causality can be brought to the interactions through considering the retarded functions, a point that  so far will not be attended here.
\begin{quote}
As a result,  the hypothesis  on the validity of $dS_4$ special relativity within the interaction domain \cite{Pereira} amounted to the {\tt prediction
of the  new phe\-no\-me\-non of con\-form\-al\-ly invariant charge neutral systems confined to clo\-sed hy\-per sphe\-ri\-cal spaces}. 
\end{quote}

The particular tangent form of the dipole potential is not universal but rather  due to the choice of the  parametrization for the second polar angle on $S^3$, namely, 
$\chi \in \left[-\frac{\pi}{2},+\frac{\pi}{2} \right]$. This choice is arbitrary and could be changed to $\chi\in \left[ 0, \pi\right]$ in which case the tangent goes into a cotangent and the sec$^2$ into csc$^2$, as originally considered  by \cite{Birgitta}. From now onward we shall switch to this very parametrization which turns out to be more suited for  physical interpretations, to be presented  in the subsequent section.

 \begin{figure}
\centering
{\includegraphics[width=7.5cm]{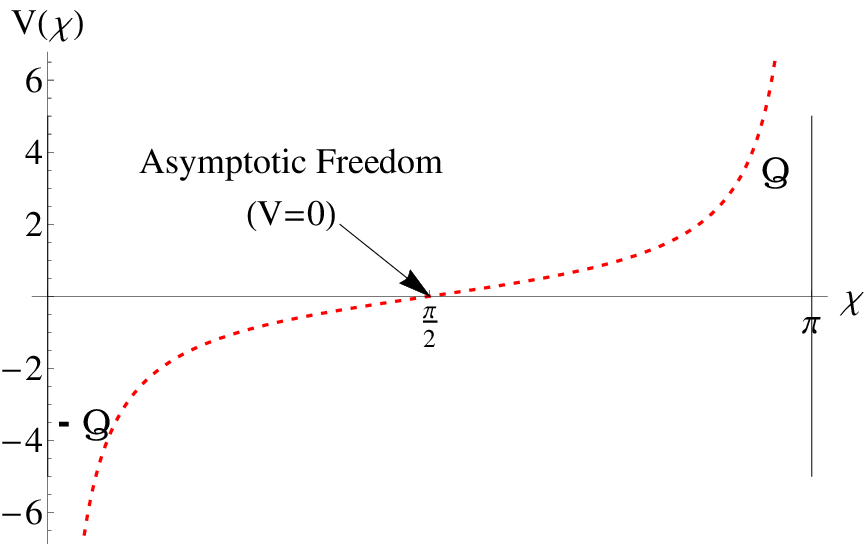}}
\caption{The conformal color dipole confining potential in the $\chi\in \left[ 0,\pi \right]$ parametrization. }
\end{figure}

Notice that ${\mathcal Q}$ stands for dimension-less charges. In terms of dimensional charges, $q$, related to $Q$ via,
\begin{equation}
{\mathcal Q}=\frac{q}{\sqrt{\hbar c}},
\label{charge_adim}
\end{equation}
the potential perceived by another charge, $q/\sqrt{\hbar c}$, is
\begin{eqnarray}
V(\chi) &=& \frac{q^2}{
4\pi {\hbar c}\epsilon_0
}\cot \chi, \quad \chi\in \left[ 0,\pi \right].
\label{Clr_Dpl}
\end{eqnarray}
{}For example, in the case of electrostatic, $q$ is taken as the electron charge, $e$, in which case the special notation of,
\begin{equation}
\alpha =\frac{e^2}{4\pi \hbar c \epsilon_0 }
\label{ED_cnst}
\end{equation}
known as the fundamental coupling constant of  Electrodynamics, is introduced.
In case the dipole is acting on a $Ze$ charge, one finds the standard Coulomb interaction, 
\begin{equation}
V^{ \left( \alpha Z\right) }_C(\chi) = 
\alpha Z\cot \chi, \quad \chi\in \left[ 0,\pi \right].
\label{Clr_Dpl_Z}
\end{equation}
In Fig.~5 we display a dipole interaction, now in the $\chi\in \left[0,\pi \right]$ parametrization. \\

\noindent
The potential in (\ref{Clr_Dpl}) can be applied as a perturbance of the free quantum motion on $S^3$ in (\ref{Scraf}), in which case the following wave equation emerges,
\begin{eqnarray}
{\mathcal H}_{\mbox{Pert}}(\chi,\theta,\varphi)
&& U_{\ell n}^{(\alpha Z)}(\chi )Y_{\ell m}(\theta,\varphi)=\frac{\hbar^2c^2}{R^2}\left({\epsilon}_{\ell n}^{(\alpha Z)}\right)^2 
U_{\ell n}^{(\alpha  Z)}(\chi ) Y_{\ell m}(\theta,\varphi),
\label{RMt_WEQ}\\
{\mathcal H}_{\mbox{Pert}}(\chi,\theta,\varphi)&= &
\left( -\frac{\hbar^2 c^2}{R^2}\frac{{\mathrm d}^2}{{\mathrm d}\chi^2} +V^{(\alpha Z)}_C (\chi)\right){\mathbf I}(\theta,\varphi),\label{Hmes}\\
V^{(\alpha Z )}_C (\chi )&=& 
V_{\mbox{Sc}}(\chi)  +V(\chi)=
\frac{\hbar^2 c^2}{R^2}\frac{\ell(\ell +1) }{\sin^2\chi } -\alpha Z \frac{\hbar^2 c^2}{R^2} \cot \chi,\nonumber\\
\label{reparm}
\end{eqnarray}
where ${\mathbf I}(\theta,\varphi)$ is the identity operator in the $(\theta,\varphi)$ variables.
Because the perturbance of the free quantum motion on $S^3$ takes place only in the $\chi$ variable, while leaving intact  the  motion in the $\theta$ and $\varphi$ variables, the ${\mathcal H}_{\mbox{Pert}}$ operator depends non-trivially only on the $\chi$ variable and is in reality one-dimensional.
The energies, $ \frac{\hbar ^2 c^2}{R^2}\left({\epsilon}_{\ell n}^{(\alpha Z)}\right)^2$, of the states bound within this potential are known \cite{Khare} and given in  as,
\begin{eqnarray} 
 \frac{\hbar ^2 c^2}{R^2}\left({\epsilon}_{\ell n}^{(\alpha Z)}\right)^2&=&
\frac{\hbar^2c^2}{R^2}\left[ -\frac{\alpha^2 Z^2}{4\left( K+1 \right)^2} -(K+1)^2\right],\nonumber\\
K=\ell +n, &\quad&  K=0,1,2,....
\label{enrg_tRM}
\end{eqnarray}
The ``spectral'' mass formula resulting then from this potential reads,
\begin{equation}
M^2=\frac{\hbar^2c^2}{R^2}\left( \epsilon_{\ell n}^{(\alpha Z )}\right)^2=\frac{\hbar^2c^2}{R^2}(K+1)^2 -\frac{\hbar^2c^2\alpha^2 Z^2}{4R^2 (K+1)^2} +c_0.
\label{mass-frla_ED}
\end{equation}
In eqs.~(\ref{RMt_WEQ})-(\ref{reparm}) one recognizes the one-dimensional wave equation with a version of the trigonometric Rosen-Morse potential,  
known from the super-symmetric quantum mechanics to be exactly solvable \cite{Khare}.

Identifying now the ``charge'' with the color charge, $\alpha$ is replaced by the QCD strong coupling, $\alpha_s$, while $Z$ is replaced by the number of colors, $N_c$ \cite{EPJA16},\cite{Addendum2016}. 
In this way, the following potential has been found,
\begin{eqnarray}
V^{(\alpha_sN_c)}(\chi)&=&\frac{\hbar^2 c^2}{R^2}\frac{\ell(\ell +1) }{\sin^2\chi } -\alpha_sN_c \frac{\hbar^2 c^2}{R^2} \cot \chi,
\label{cmplt_ptn}
\end{eqnarray} 
to be relevant in quark models.
Correspondingly, the mass formula emerges as,
\begin{equation}
M^2=\frac{\hbar^2c^2}{R^2}\left( \epsilon_{\ell n}^{(\alpha_s N_c )}\right)^2=\frac{\hbar^2c^2}{R^2}(K+1)^2 -\frac{\hbar^2c^2\alpha_s^2N_c^2}{4R^2 (K+1)^2} +c_0.
\label{mass-frla}
\end{equation}
To recapitulate, the result is that in hypothesizing validity of  $dS_4$ special relativity \cite{Pereira}, it became possible to predict the form of a potential generated by  equal numbers of charges and anti-charges locked (confined) on a closed hyper-spherical space, much alike the phenomenon of color confinement in QCD.

\section{The experimental verification }
The phenomenon of charge confinement predicted by the $dS_4$ special relativity in the previous section is well known from the theory of
strong interactions, the Quantum Chromodynamics (QCD) \cite{textbook}, in which the fundamental matter degrees of freedom, the quarks, are endowed with so called ``color'' charges. There are three color charges, and three color anti-charges, although the 
strongly interacting particles, the  hadrons, are by themselves all color charge neutral. For example, the hadrons of integer spin, the mesons, are constituted by effective degrees of freedom, termed to as ``constituent''  quarks, 
and ``anti-quarks'',  which are equipped by opposite color charges. 
Within this context it is legitimate to ask the question as to what extent the scenario developed in  the preceding  sections and especially of the previous section 5 could apply to the description of the experimentally reported  meson masses.
The question we are posing  is as to what extent the dual ``trajectory''- and ``spectral'' mass formulas in the respective eqs.~(\ref{My_trjct}), and (\ref{InvMass_ReE2}) 
are suited for the description of the meson mass dependencies on their angular momenta. Such an analysis has been performed in \cite{EPJA16} and we here limit ourselves to only briefly highlight the results.\\

\noindent
In \cite{EPJA16} we analyzed 71 reported unflavored mesons belonging to the four families of the $f_0$, $a_0$, $\pi$, and $\eta $ mesons.
Data convincingly confirmed   the conformal hydrogen-like $(K+1)^2$-fold degeneracy of the levels in accord with (\ref{mass-frla}). In addition, their splittings followed pretty much those of the states bound within the trigonometric Scarf potential in (\ref{InvMass_ReE2}). Also  the dual mass formula in (\ref{My_trjct}) turned out  well applicable. The only splittings which could  not come out well from the formulas under discussions were the splittings between the ground states and the first excited states. They could be accounted for by the help of the extension in (\ref{cmplt_ptn}). \\

\noindent
Usage of (\ref{Clr_Dpl}) as a perturbance was justified by the suggestion to consider mesons as made of two types of color-dipoles, one constituted by a  quark and an anti-quark, and the other by a gluon and an anti-gluon. The meson Hamiltonian,
${\mathcal H}_{\mbox{Pert}}$, of such a system, given in (\ref{Hmes}), can then  be approximately separated into a dominant  free 4D rotational motion of say, the supposedly light quark-anti-quark pair, described by the Hamiltonian ${\mathcal H}_{\mbox{Sc}}$ in (\ref{paso11}), perturbed by  the ``residual'' color-dipole interaction, $V_{CD}(\chi)$ in (\ref{Clr_Dpl}), generated by a background gloun-antigluon dipole.
Such a picture is consistent with the idea of the constituent quarks as effective degrees of freedom, i.e. as fundamental quarks ``dressed'' by gluons.
In adopting ${\mathcal H}_{\mbox{Pert}}$ as a working hypothesis, one is of course neglecting the perturbance of the glueball by the color-dipole potential generated by the quark-anti-quark pair, and the tensor interaction between the two dipoles \cite{Gaillol}.

Notice that the interpretation of hadrons as states bound within a well potential is adequate near the rest frame, while their interpretation as resonances transmitted through a barrier is suited for high to  ultra-relativistic energies.
The rest frame is acceptable as an approximation at low transferred momenta in 
the so called infrared (IR) regime of QCD, while the adequate tool for hadron description at high-and ultra-high transferred momenta,i.e. close to the ultraviolet (UV) regime of QCD, is provided by the scattering matrix. Stated differently,  within the QCD context, the quantum motions on the open time-like geodesics 
can be associated with  processes approaching  the ultraviolet regime, while those on the closed space-like geodesics can be associated with near rest frame processes, i.e. with the infrared regime of QCD. In this sense, the achievement of the model advocated  in \cite{EPJA16} is to have
\begin{itemize}

\item associated (to a good approximation) each one of  the infrared and the ultraviolet regimes of QCD with free motions along the respective closed hyper-spherical space-like --, and the open hyperbolic time-like geodesics on a conformally symmetric  $dS_4$ space-time,

\item established with respect to meson description a duality between
states bound within a well-(IR), and resonances transmitted through a barrier (UV) potentials,

\item  gained the new understanding about the inevitable color-neutrality of hadrons in the infrared as a consequence of the innate charge-neutrality of the closed hyper-spherical space-like $dS_4$ geodesics and predicted a confinement phenomenon known in QCD,

\item revealed relevance of  conformal symmetry  in both regimes of QCD.

\end{itemize}
 Notice that color-neutrality in the ultraviolet occurs by virtue of the dual meson description and is not inevitable there  because on open space-times free charges and therefore a deconfined phase, are allowed to exist. 

Finally, it needs to be emphasized that the infrared  $\left (- {\mathcal Q}\right){\mathcal G}_{\frac{\pi}{2}} (\chi)$ single-color potential in (\ref{VSP}) relates by an argument shift toward, $\chi=\chi^\prime  +\pi/2$, followed by  the complexification,
$\chi^\prime \to i\chi^\prime \to \rho$, to  the expression $\left( -{\mathcal Q}\right) (-i\chi^\prime)  \cot i\chi^\prime \to - \left( -{\mathcal Q}\right) \rho \coth\rho$, 
giving  a single-color potential generated in the ultraviolet  by a soft-gluon emission of a fast quark,  stopped at a cusp \cite{Belitsky}, and in support of the IR-UV duality advocated here and in \cite{EPJA16}.
The Wilson loop calculation of such a potential, in making use of radial quantization, is practically equivalent to our calculation in section 5.

\subsection{The spectra of the $a_1$ and $f_1$ mesons}
As already mentioned above, the fundamental matter  degrees of freedom in the  theory of strong interaction, the Quantum Chromodynamics,
the quarks, can carry three different color charges. The color charge, denoted by $g(Q^2)$, depends on the square, $q^2$,  of the transferred space-like four-momentum,  via $Q^2=-q^2$, and the related fundamental coupling in QCD, $\alpha_s(Q^2)$, defined in analogy to (\ref{ED_cnst}) as,  
\begin{eqnarray}
\alpha_s(Q^2)=\frac{g^2(Q^2)}{4\pi \hbar c}, \quad Q^2=-q^2\geq 0,
\label{strong_CplCst}
\end{eqnarray}
changes with $Q^2$, a reason for which it is is termed to as ``running''.

In the following  we  complement  the analyzes of the mass distributions of the four meson families previously studied in  \cite{EPJA16} (71 masses) by  two more meson families, those of the  $f_1$, and $a_1$ mesons,  which contribute 18 more masses.
Afterwards we  extract the corresponding  $\alpha_s$ values.
Compared to the $f_0$, $\eta$, $\pi$, $a_0$ mesons, analyzed in \cite{EPJA16}, less is known about the $f_1$ and $a_1$ mesons.
The compiled data on the last two families are presented in Fig.~5. It needs to be said that only the $a_1(1260)$, $a_1(1640)$, and $f_1(1285)$
mesons are contained in the Summary Table of the Meson Particle Listings in \cite{PART} meaning that their existence is considered as fully reliable. 
All the other particles in Fig.~5 are either  included in the Meson Particle Listings, or
into the list of the less reliably established particles referred to as  `` Further States''. We here favored the $f_1(1510)$ meson as the first excited state of $f_1(1285)$ over the  $f_1(1420)$ state from the Summary Table. In this way we are suggesting that the internal $f_1(1510)$ structure  may be  closer to 
a quark--anti-quark color dipole, perturbed by the dipole potential due to  gluon--anti-gluon dipole, than the $f_1(1420)$ internal structure, which may contain strange-anti-strange quarks.  
Our point is that our conformal symmetry classification scheme, in needing  states omitted from the Summary Table,  provides an additional legitimization to their observation and consolidates their status.
Same as in \cite{EPJA16}, we here employ the mass formula in (\ref{mass-frla}) to perform a least square fit to  the meson mass distributions in Fig.~5.
The results on the three parameters, the hyper-sphere radius,  $R$, and the two dimensionless parameters $b$, and $c_0$ in (\ref{mass-frla}),  are listed in the  columns second to fourth of  Table 1. The fifth column contains the average deviation values of the corresponding fits, while in the last column we present the values of the strong coupling constant, $\alpha_s$ of QCD (normalized to $\pi$) which we  extracted by virtue of  the formula (\ref{mass-frla}). The Table shows that the two $\alpha_s/\pi$ values, in being of the order of $ \alpha_s/\pi \sim 0.80$,  remain significantly less than $1$, and  in complete agreement with  data reported in \cite{Andre}, whose  
principle  result is  that  the strong coupling constant tends to approach a fixed value in the infrared, thus opening up a so called ``conformal window'' and thereby hints on relevance of conformal symmetry in the infrared.

\begin{table} 
 \begin{center}
     \resizebox{0.85\textwidth}{!}
{\begin{minipage}{\textwidth}
\begin{tabular}{|c|c|c|c|c|c|}  
\hline 
trajectory &           R         & b     &   c$_0$ &  $\sigma^2$ [GeV]$^4$                   & $\frac{\alpha_s}{\pi}$ \\ 
\hline \hline 
\hline
$f_1$ for  $K\geq 0$ &0.58 fm     & 3.788 &5.125    &    0.1652839 GeV$^4$                      & 0.80 \\
\hline 
$a_1$ for   $K\geq 0$ &0.58 fm    & 3.726 & 5.071   &    0.09191531 GeV$^4$                      & 0.79\\
\hline\hline
\end{tabular} 
\caption{Parameters of the least square fit to the 18 masses of the $a_1$ and $f_1$ mesons, distributed over the two trajectories indicated in the first column, and  calculated employing  the mass formula in (\ref{mass-frla}). The last column contains the value of the extracted strong coupling constant, $\alpha_s/\pi$, identified by the equation (\ref{mass-frla}), with $2b=\alpha_sN_c$.  The explicit expression of $\sigma^2$ reads
$\sigma^2=\frac{\left(\sum \Delta M^2\right)^2}{N-1}$, where $\Delta M^2$ stands for the deviation of the squared experimental from the squared theoretical masses, and $N$ is the number of states in a family.
} 
\end{minipage}}
  \end{center}
\label{Table1} 
\end{table}

\begin{figure}
\centering
{\includegraphics[width=7.5cm]{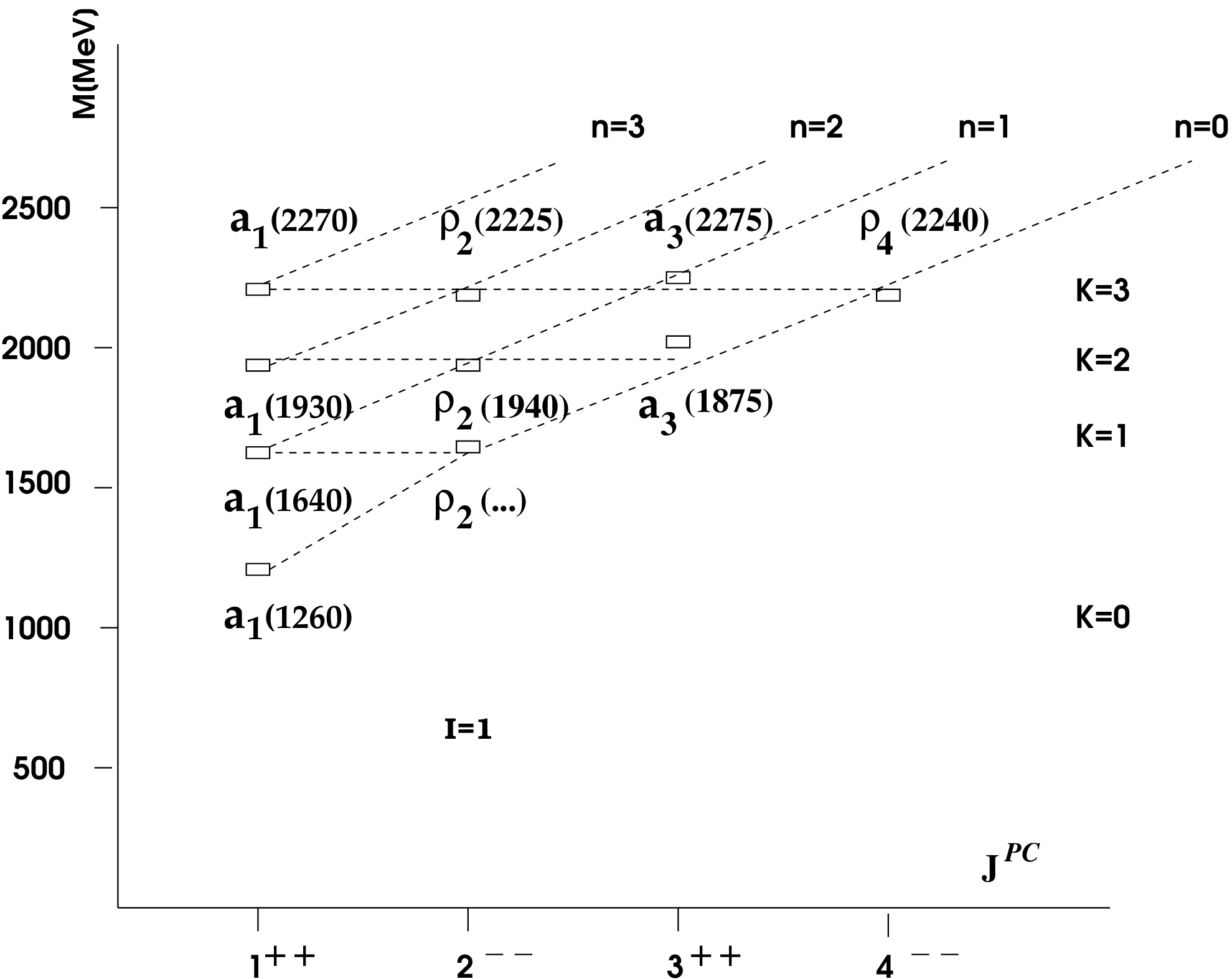}}
{\includegraphics[width=6.95cm]{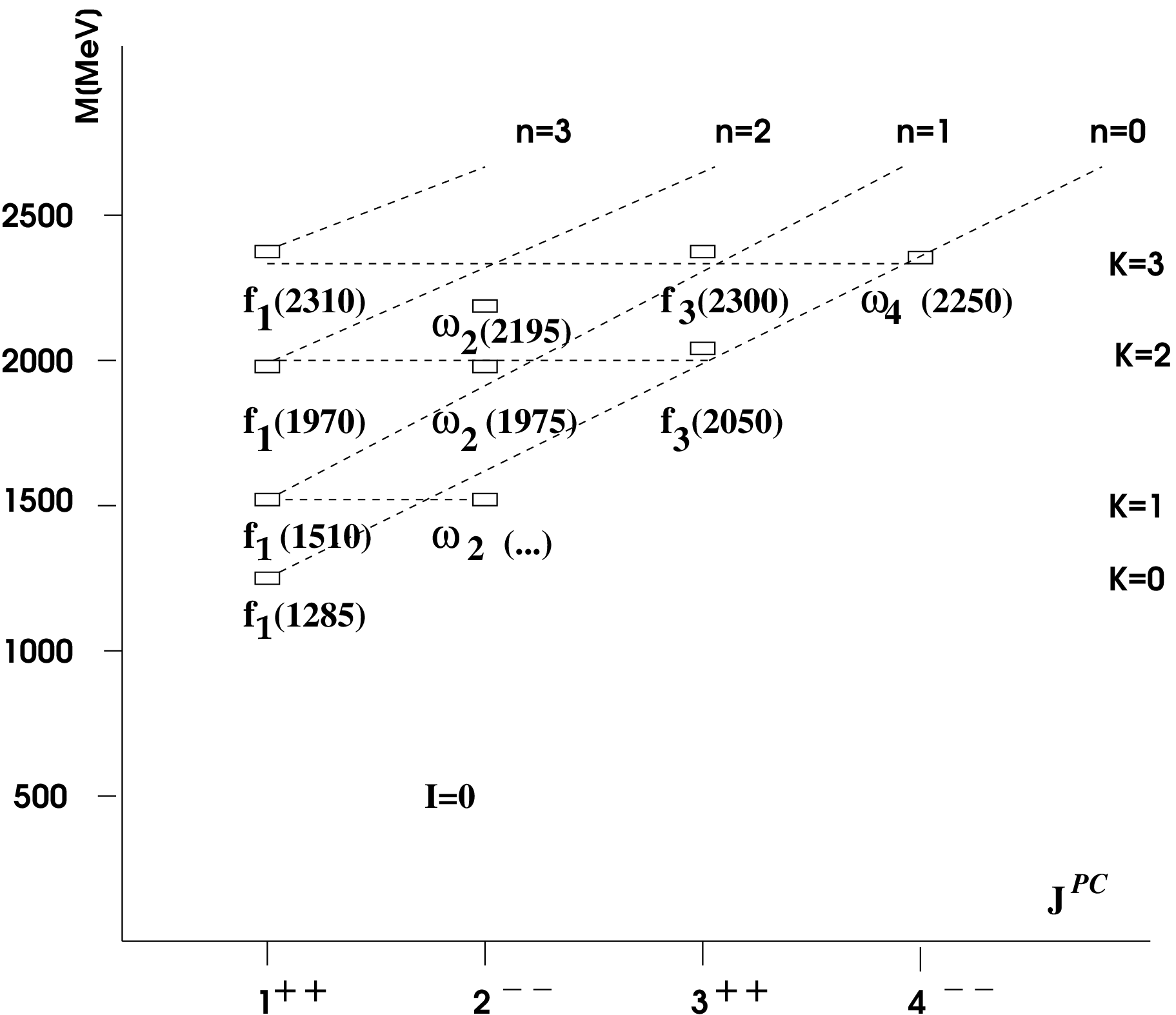}}
\caption{The excitations of the isovector  $a_1(1260)$ meson (left) and the isoscalar $f_1(1285)$ meson (right). ``Missing'' states are denoted by (...).
Here, $J$ is the total angular momentum value of the quark-diquark system obtained from coupling the total spin $S=1$, to the relative angular momentum, $\ell=0,1,2,3$, and $CP$ give in their turn the $C$ and $P$ quantum numbers.
Recall the definition of $K=n+\ell$ in (\ref{My_trjct}). }
\end{figure}

\subsection{The electric form factor of the pion}

In our previous work \cite{NPhA18} we formulated on $S^3$ the Dirac equation with the color-confining dipole potential in (\ref{Clr_Dpl}), found the spinor solutions in closed forms, and used them in the calculations of the proton's and neutron's  electric-charge, $G_{p/n}^E(Q^2)$, and magnetic dipole, $G_{p/n}^M(Q^2)$ form factors, together with the respective Dirac- and Pauli form factors, 
$F_1^{p/n}(Q^2)$, and $F^{p/n}_2(Q^2)$.

The electric charge form factor of the proton is determined by the Fourier transform of the electric-charge density, $\rho^p ({\vec r}\, )$,  as
\begin{eqnarray}
G_p^E(Q^2)&=&\int_{V( {\vec r}\,)\to \infty} 
\rho^p(r)
e^{i\vec q\, \cdot \vec r } d^3{\vec r}, \quad \rho^p({\vec r}\, ) =
|\Psi_{gst}({\vec r}\, )|^2,\\
Q^2&=&-q^2,
\end{eqnarray}
where $Q^2$ equals the negative squared of the transferred space-like momentum, $q^2<0$,  $V({\vec r}\, )$ is the space integration volume, and $\Psi_{gst} ({\vec r}\, )$ is the Dirac spinor of ground state. 
The proton's magnetic dipole form-factor is given by the Fourier transform of the magnetic dipole density, $\rho^p_{mag}({\vec r}\, )$ as,
\begin{equation}
G_p^{M}(Q^2)=\int_{V({\vec r}\,)\to \infty} \rho^p_{mag}(r)e^{i \vec q\, \cdot \vec r } 
d^3 {\vec r}, \quad \rho^p_{mag} ({\vec r}) ={\bar \Psi}_{gst}({\vec r}\, )
\gamma_5  \Psi_{gst}({\vec r}\, ).
\end{equation} 
The relation between $G_p^E(Q^2)$, and $G_p^M(Q^2)$ to the Dirac and Pauli form factors, $F_1^p(Q^2)$, and $F_2^p(Q^2)$, is given by,
\begin{eqnarray}
F_1^p(Q^2)&=&\frac{G_E^p(Q^2) +\tau_p G_M^p(Q^2)}{1+\tau_p},\quad \tau_p =\frac{Q^2}{2M_p^2},\nonumber\\
\kappa_p F^p_2(Q^2)&=&\frac{G_M^p(Q^2) -G_E^p(Q^2)}{1+\tau_p}, \quad \kappa_p =\mu_p-1,
\end{eqnarray}
where $\mu_p$ is the proton's magnetic dipole moment.
The neutron form factors are expressed by means of the proton's form factors as,
\begin{eqnarray}
G_n^E(Q^2)&=&-\frac{\mu_n\tau_n}{1+B\tau_n}G_p^E(Q^2),\quad \tau_n =\frac{Q^2}{2M_n^2}\nonumber\\
\frac{G_M^n(Q^2)}{\mu_n}&=&\frac{G^p_M(Q^2)}{\mu_p},
\end{eqnarray}
where $B$ is a parameter, while $\mu_n$ is the magnetic dipole moment  of the neutron.  In \cite{NPhA18} expressions in closed form for all the above form-factors have been reported, which we opted not to repeat here for the sake of not overloading the presentation.
On the other side, all the form factors can also be alternatively expressed in terms of  the respective form factors of the constituent $u$ and $d$ quarks. Specifically for the Dirac and Pauli form factors the following expressions hold valid,

\begin{eqnarray}
F^u_i(Q^2)&=& 2F^p_i(Q^2)+F^n_{i}(Q^2), \quad i=1,2,
\label{F12u}\\
F_{i}^d(Q^2)&=& F_i^p(Q^2)+2F^n_i(Q^2).\label{Fd12}
\end{eqnarray}

These form factors, extracted from the $F^{1}_p(Q^2)$ and $F^2_n(Q^2)$ expressions reported in \cite{NPhA18} are plotted in Fig.~6. 

\begin{figure}
\centering
{\includegraphics[width=7.5cm]{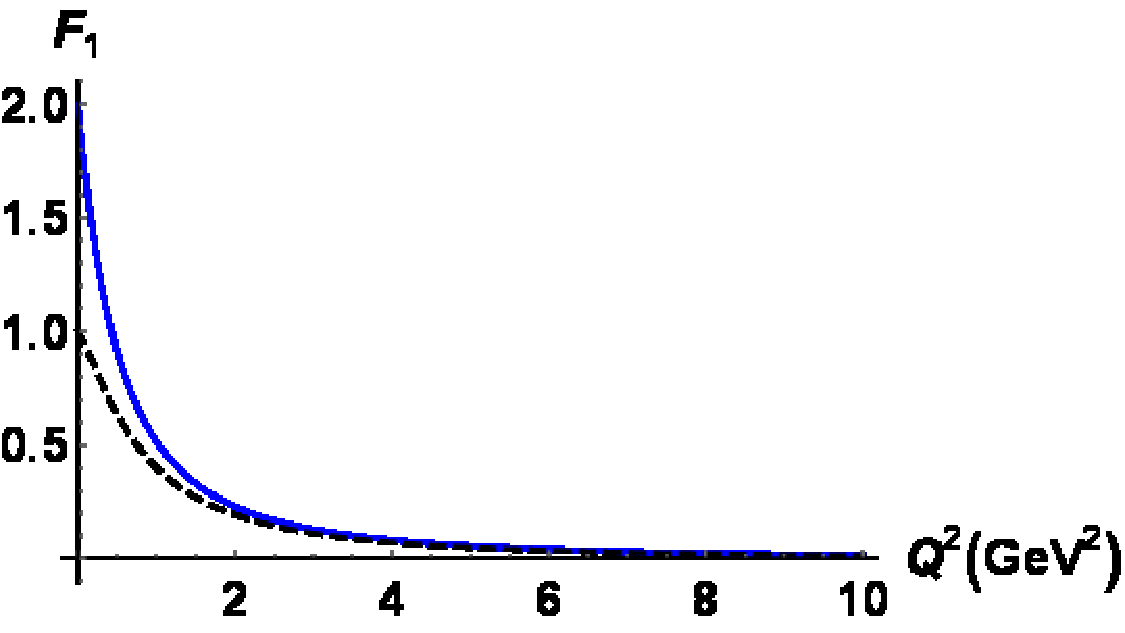}}
{\includegraphics[width=7.5cm]{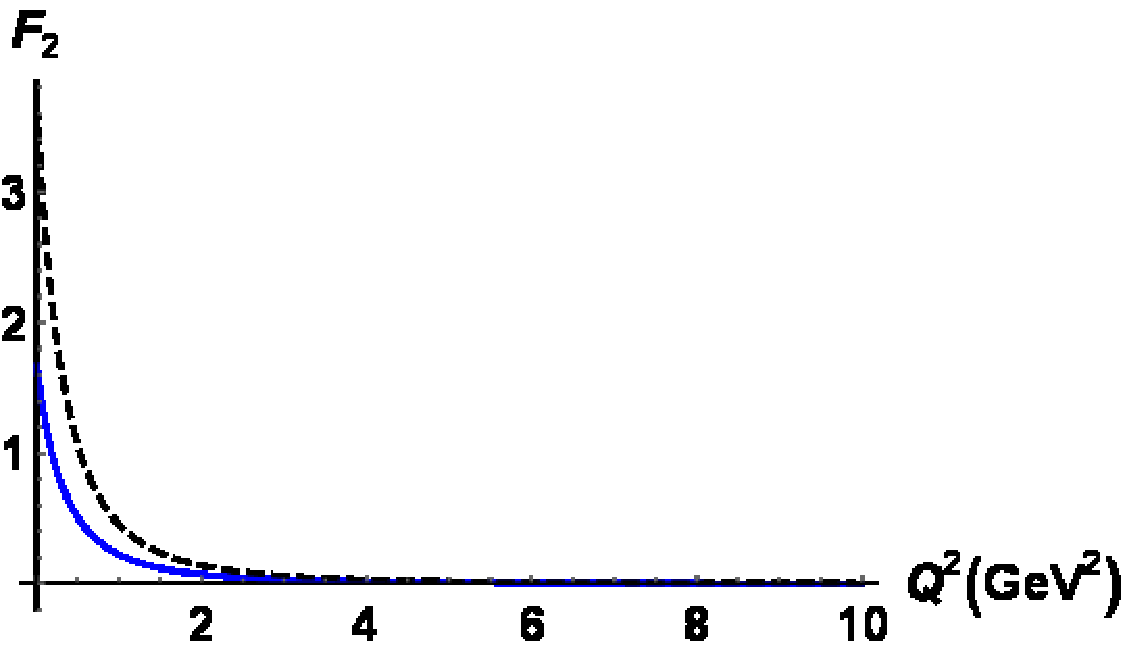}}
{\includegraphics[width=7.5cm]{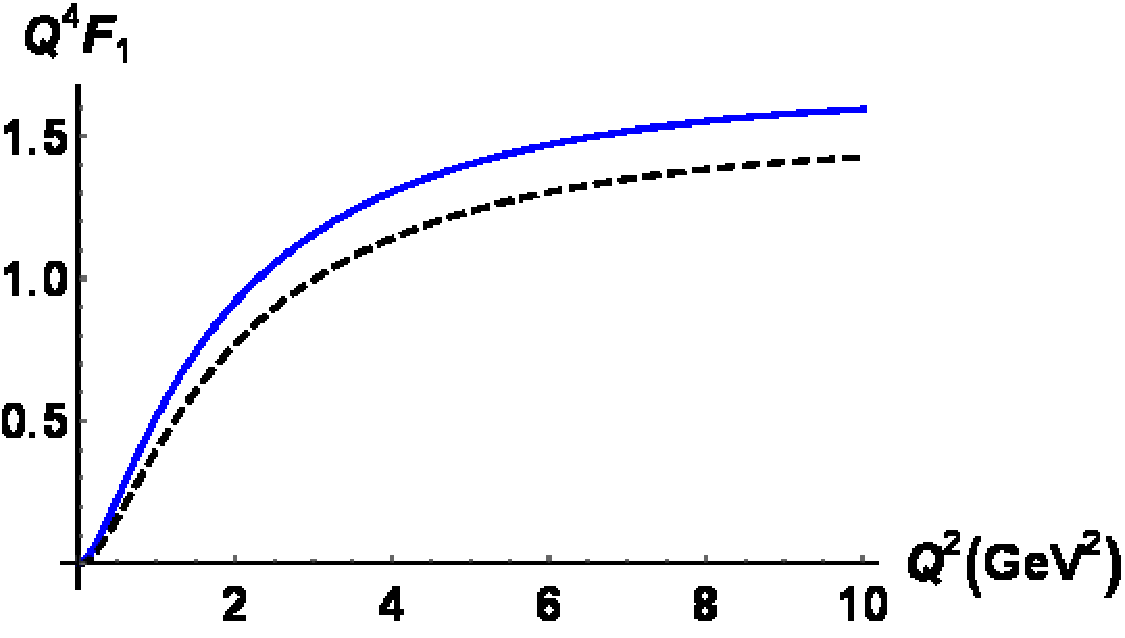}}
{\includegraphics[width=7.5cm]{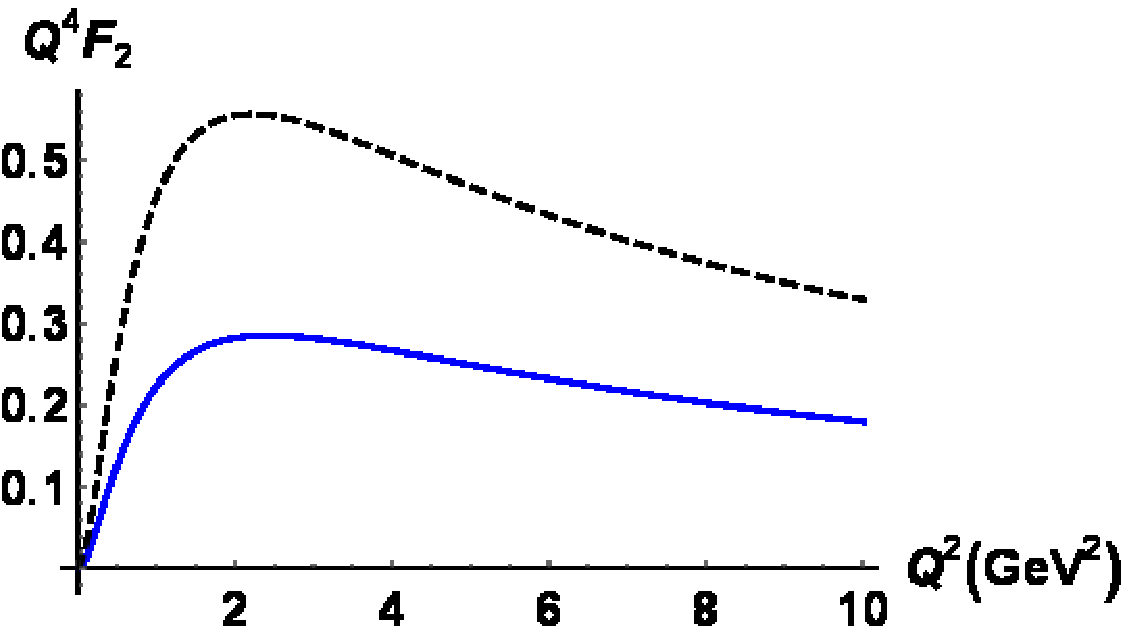}}
\caption{The Dirac  and Pauli form factors of the $u$ (solid lines) and $d$ (dashed lines) quarks (left and right top figures), and their respective scalings (left and right bottom figures).}
\end{figure}
We observe that our predictions on the light quark form factors  are of same quality  
as those  reported  in \cite{Dipankar} and worked out 
within the framework of the related $ADS_5/CFT_4$ formalism.
Knowing the quark form factors allows one to obtain the  form factor $F_\pi(Q^2)$ of the positively charge pion,  $\pi^+$, as
\begin{eqnarray}
F_\pi(Q^2)=\frac{1}{2}\left[ \frac{2}{3}F_1 ^u(Q^2)\right]  - \left[ -\frac{1}{3}F_1^d(Q^2)\right],
\label{piFF}
\end{eqnarray}
where we maintained same potential parameters as those fitted to the proton's and neutron's form factors.

Here, the last term on the r.h.s. in the latter equation stands for  the $\bar d$ form factor. 
The satisfactory result, displayed in Fig.~7 together with its  comparison with data taken from \cite{Almd},
provides a further support for the convenience of our suggested approach to hadron spectroscopy.   

\begin{figure}
\centering
{\includegraphics[width=7.5cm]{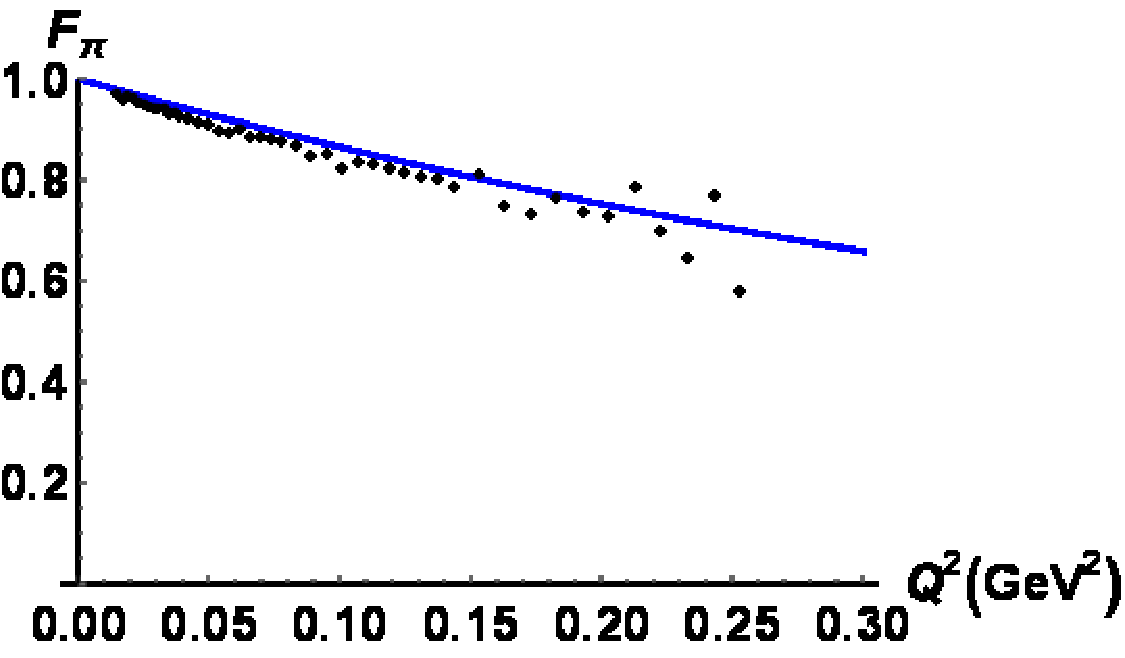}}
\caption{The pion electric form factor, $F_\pi(Q^2)$, calculated with the aid in (\ref{piFF}) (solid line) in comparison to data taken form \cite{Almd} (dots).}
\end{figure}

\section{Conclusions}
Our analyzes show that the experimentally reported conformal symmetry patterns of the light flavored meson mass distributions  are well modelled by  
 a conformal $dS_4$  internal  space-time, considered as located  outside of the Minkowskian light  cones attached to local observes on this space, and hypothesized as the geometry relevant for the description of the virtual degrees of freedom inside hadrons. 
Along this line, the interaction in the infrared regime of QCD among the virtual hadron constituents has been defined by the Laplacian of the closed  space-like hyper-spherical geodesics $S^3$ on  $dS_4$,  where quark and gluon systems are inevitably and necessarily colorless,
In effect,  hadrons in the infrared have been viewed as color-anti-color dipoles in free conformal four-dimensional quantum rotation, described by the $\ell (\ell +1)\mbox{csc}^2\chi$ potential, or, rotation  slightly perturbed by the also conformal color-dipole  potential, $(-\alpha_sN_c\cot\chi )$, attributed to a gluon-anti-gluon color dipole. The magnitude of the perturbance is proportional to the product of the number of colors, $N_c$ by the strong coupling constant, $\alpha_s$, by virtue of  (\ref{cmplt_ptn}), a circumstance that allowed us to extract from data on  meson masses reasonable  $\alpha_s$-values, and to conclude on the relevance of a dynamical  conformal symmetry in both the extreme regimes of QCD.

Thus, the dynamical conformal symmetry, via $dS_4$ special relativity, backs up the internal color gauge group, $SU(3)_C/Z_3$, in QCD. However, on  open spaces, be them the open time-like geodesics on $dS_4$, or its casual patches, on which no charge-neutrality is required, color excess could exist within this scenario and become detectable.
Our findings favor a  space-time, with one extra (infinite) dimension in the 
internal coordinate space.
The scheme is  Fourier transformable and can be re-formulated in momentum space.
Models of this type are free from infrared divergences \cite{Belitsky}, a reason for which  considering QCD on closed spaces is appealing (also see \cite{Kharzeev} for further reading).
To the best of our knowledge, the model under discussion is so far the only one which allows to estimate the fundamental coupling constant of QCD from analyzes of global (integral) properties of hadron systems such as spectra.
Usually, these values are extracted from data on the spin structure function \cite{Andre}. Our major conclusion is that behind the curtains of conformal symmetry and color-confinement in QCD one may encounter the deSitter $dS_4$ special relativity as a string-puller.
Finally, the quality reached in the description of the electromagnetic form factor of the pion indicates that the $u$  quark in the pion is same as in the proton, while the $\bar d$ quark in the pion is related to the $d$ quark in the proton by charge conjugation.


\begin{thebibliography}{9}
\bibitem{PART} K.\ A.\ Olive {\it et al.} ( Particle Data Group), Review of Particle Physics,
Chinese Physics C {\bf 38} 090001 (2014) and 2015 update.
\bibitem{Rayzuddin}Fayyazuddin and Riazuddin, {\it A Modern Introduction to Particle Physics}, 2nd edition  (World Scientific, Singapore, 2000)
\bibitem{MIT}A.\  Chodos, R.\ L.\ Jaffe, K.\ Johnson, and C.\ B.\ Thron, Phys.\ Rev.\ D {\bf 10} (1974) 2599. 
\bibitem{Kharzeev}D.\ Kharzeev, E.\ Levin, and K.\ Tuchin,
Phys.\ Rev.\  D {\bf 70} (2004) 054005 .
\bibitem{EPJA16} M.\ Kirchbach and  C.\ B.\ Compean, Eur.Phys.J. A  {\bf 52}  (2016) 210.
\bibitem{Pereira} R.\ Aldrovandi, J.\ P.\ Beltr\'an Almeida, and J.\ G.\ Pereira,
Class. Quant. Grav. {\bf 24} (2007)  1385.

\bibitem{Kelogg} O.\ D.\ Kelogg, {\it Foundations of Potential Theory} (Dover, New York, 1953). 

\bibitem{Addendum2016} M.\ Kirchbach and  C.\ B.\ Compean, Eur.Phys.J. A  
{\bf 53}  (2017) 65.

\bibitem{Barut} A.\ O.\ Barut and R.\ Wilson, Phys.\ Lett.\ A {\bf 110} (1983) 351.
\bibitem{NPhA18} M.\ Kirchbach and  C.\ B.\ Compean, Nucl.\ Phys.\  A  {\bf 980}  (2018) 32.


\bibitem{Slava} Slava Rychkov, {\it EPFL Lectures on Conformal Field Theory in $D\geq 3$ dimensions}, E-print arXiv:1601.05000[hep-th]. 


\bibitem{Coreans} Yoonbai Kim, Chae Young Oh, and Namil Park, J.\  Korean Phys.\ Soc.\ {\bf 42} (2003) 573 .



\bibitem{Khare} F.\ Cooper, A.\ Khare, and  U.\ P.\ Sukhatme,
{\it Supersymmetry in Quantum Mechanics} (World Scientific, Singapore, 2001).
\bibitem{Prestley} Andrew Pressley, {\it Elementary Differential Geometry} (Springer, London Dordrecht Heidelberg New York, 2012).

\bibitem{Cevik}   D.\ Cevik, M.\ Gadella, S.\ Kuru, and  J.\ Negro,
Phys.\ Lett.\ A {\bf 380} (2016) 1600.


\bibitem{ED_CS}  H.\ Bateman, 
Proc.\ London Math.\ Soc.\ (ser. 2) {\bf 7} (1909)  70.

\bibitem{LandLif}L.\ D.\  Landau and  E.\ M.\ Lifschitz, {\it The Classical  Theory of Fields}, Vol. 2 of A Course of Theoretical Physics,
 3d edition (Pergamon Press 1971) p.335. 
\bibitem{PouriaPedram}Pouria Pedram,  Am.\ J.\ Phys.\ {\bf 78}  (2010) 403.
\bibitem{Fubini} S.\ Fubini, A.\ J.\ Hanson, and R.\ Jackiw, Phys.\ Rev.\ D {\bf 7} (1972) 1732.
\bibitem{Birgitta}B.\ Alertz, Ann.Inst.Henri Poincar\'e, {\bf 53}  (1990) 319.
\bibitem{textbook} Kerson Huang, {\it Quarks, Leptons and Gauge Fields} (World Scientific, Singapore, 1982).
\bibitem{Gaillol}J.-M.\ Gaillol and  M.\ Trulsson, J.\ Chem.\ Phys. {\bf 141} (2014) 124111. 
\bibitem{Belitsky}  A.\ V.\ Belitsky, A.\ S.\ Gorsky, and G.\ P.\ Korchglemsky,
Nucl.\ Phys.\ B {\bf 67} (2003) 3.



\bibitem{Andre} A.\ Deur, V.\  Burkert, J.\ P.\  Chen, and  W.\ Korsch,
Phys.\ Lett. {\bf 665} (2008)  349.


\bibitem{Dipankar} Dipankar Chakrabart, Chandan Mondal, Eur.\ Phys.\ J.\ C {\bf 73} (2013) 2671.


\bibitem{Almd} S.\ R.\ Amendolia et al., Nucl.\ Phys.\ B {277} (1986) 168.

\end{thebibliography}
\end{document}